\documentclass[pdflatex]{scrartcl}
\usepackage[amsthm,thmmarks]{ntheorem}
\usepackage{amsmath}
\usepackage{amssymb}
\usepackage{mathrsfs}
\usepackage{bm}
\usepackage{bbm, dsfont}
\usepackage[T1]{fontenc}
\usepackage{pxfonts}
\usepackage{comment}
\usepackage{url}
\usepackage{graphicx}
\usepackage{float}
\usepackage{wrapfig}
\usepackage{booktabs}
\usepackage{multicol}
\usepackage{multirow}
\theoremstyle{break}

\usepackage{authblk}

\usepackage[backend=biber, firstinits=true, sorting=none, style=ieee-alphabetic, doi=false,isbn=false,url=false, eprint=false]{biblatex}
\AtEveryBibitem{\clearfield{month}}
\AtEveryCitekey{\clearfield{month}}

\title{Continuous-time Portfolio Optimization for Absolute Return Funds}
\date{}
\author[$\dag$]{Masashi IEDA\thanks{e-mail: ieda[at]rs.tus.ac.jp}}
\affil[$\dag$]{
	Department of Business Economics, School of Management \authorcr
    Tokyo University of Science  \authorcr
	1-11-2 Fujimi, Chiyoda-ku, Tokyo, 102-0071, Japan
}

\begin{document}

\maketitle

\vspace*{-2em}
\begin{center}
	\textbf{Abstract}
\end{center}

\begin{abstract}
\noindent
This paper investigates a continuous-time portfolio optimization problem with the following features:
(i) a no-short selling constraint;
(ii) a leverage constraint, that is, an upper limit for the sum of portfolio weights; and
(iii) a performance criterion based on the lower mean square error between the investor's wealth and a predetermined target wealth level.
Since the target level is defined by a deterministic function independent of market indices, it corresponds to the criterion of absolute return funds.
The model is formulated using the stochastic control framework with explicit boundary conditions.
The corresponding Hamilton-Jacobi-Bellman equation is solved numerically using the kernel-based collocation method.
However, a straightforward implementation does not offer a stable and acceptable investment strategy; thus, some techniques to address this shortcoming are proposed.
By applying the proposed methodology, two numerical results are obtained: one uses artificial data, and the other uses empirical data from Japanese organizations.
There are two implications from the first result: how to stabilize the numerical solution, and a technique to circumvent the plummeting achievement rate close to the terminal time.
The second result implies that leverage is inevitable to achieve the target level in the setting discussed in this paper.
\end{abstract}


\pagebreak
\section{Introduction}
Portfolio optimization is one of the challenging problems for both academic researchers and practitioners.
Here, we focus on the investment strategy of long-term investors such as pension funds and insurance companies.
Naturally, they change their decision from time to time according to market conditions.
Hence, a single-period model in which investors can decide only at the beginning appears to be a subordinate choice for them.
A continuous-time model is a common approach to this issue, and thus, has been investigated intensively by academic researchers.
The review paper by \autocite{Sundaresan2000} provides an overview of studies
 from 1969, when the seminal paper by Merton \autocite{Merton1969} was published, until the end of the twentieth century.

We briefly mention more recent studies related to continuous-time portfolio optimization.
(i) Regarding objective functions, we list some studies:
the risk-sensitive criterion \autocite{Hata2017};
the conditional value-at-risk (or expected shortfall) \autocite{Miller2017};
the power utility \autocite{Elie2008}; and
the exponential utility \autocite{Ma2019}.
(ii) The mean-variance portfolio optimization has been also investigated in the continuous-time setting.
The difficulty in this approach is that the control problem describing mean-variance criteria does not follow the standard dynamic programming principle.
See e.g., \autocite{Li2002}, \autocite{WANG2012}, \autocite{Ismail2019} and \autocite{DeFranco2018a} for more details.
(iii) Economic conditions dynamically change from time to time and we cannot discern them ex-ante.
Therefore, the model parameter uncertainty is more critical in continuous and multi-period models compared to the single-period model.
To circumvent this issue, the following studies address the robustness of portfolio optimization: \autocite{Xidonas2020}, \autocite{Huang2008}, \autocite{Cong2017} and \autocite{FORSYTH2017}.

The continuous-time model is preferable for long-term investors; however, it does not appear to be widely applied by practitioners.
One of the practical shortcomings of this approach is that the obtained optimal portfolio weights tend to contain large values.
The same problem occurs in the single-period mean-variance framework.
Practitioners usually circumvent this issue by imposing various constraints, although their effectiveness is debatable.
The main focus of previous studies on the continuous-time model is to obtain analytical solutions.
However, the analytical solution for the constrained model is rarely available and the literature dealing with this constraint is limited (e.g., \autocite{Li2002} and \autocite{Gao2017}).

In this paper, we investigate a continuous-time portfolio optimization problem with
 (i) a no-short selling constraint;
  (ii) a leverage constraint, that is, an upper limit for the sum of portfolio weights;
   and (iii) a performance criterion based on the lower mean square error (LMSE) between an investor's wealth and a predetermined target wealth level.
Constraints (i) and (ii) are the typical requirements of institutional investors.
Criterion (iii) improves the mean square error frequently employed in previous works in the sense that we are able to avoid the unfavorable penalty which occurs when the investor's wealth exceeds the target level.
Note that the target wealth level is defined as a deterministic function of time, and thus, is independent of any future values, such as returns of market indices.
Hence, it corresponds to the criterion of absolute return funds.
The target level in this paper is almost characterized by the required absolute return.

We contribute to the literature by providing a concrete methodology to obtain an acceptable investment strategy for practitioners.
This is accomplished by developing the model and methodology discussed in previous works \autocite{Ieda2014}.
Under the leverage constraints, the quadratic approximation method used in \autocite{Ieda2014} only provides a proxy for the solution of the Hamilton-Jacobi-Bellman (HJB) equation representing the heart of the problem.
It further restricts the formulation of the leverage constraints in an unnatural way.
The former issue can be resolved by employing the kernel-based collocation method (see \autocite{Kansa1990} and \autocite{Nakano2014}) instead of the quadratic approximation method.
However, we cannot obtain a stable and acceptable solution by simply applying the kernel-based collocation method.
We present some techniques to resolve this, such as expanding the computational domain or introducing the margin rate.
Thanks to the new approximation method, the latter point is also resolved.
We can reformulate the model, including the constraints, more naturally.
Moreover, the dimension of the state variable decreases from two to one.
This significantly reduces the computational cost.
In the new formulation, we find the explicit boundary condition, in contrast to what is unknown in the previous work.
Although the numerical method employed here can be applied without boundary conditions, it improves the accuracy of the numerical solution.

The remainder of this paper is organized as follows.
Section \ref{sec:model_formulation} introduces our model.
We formulate the investment problem in the continuous-time stochastic control framework.
The goal is to obtain the corresponding HJB equation with explicit boundary conditions.
In Section \ref{sec:num_method}, we present a numerical method for solving the HJB equation.
We outline a concrete procedure to obtain the numerical solution and describe a technique for stabilizing the solution.
The numerical results are presented in section \ref{sec:num_results}.
The results with artificial data show features of the numerical solution, and some implications for improving the investment are provided.
Incorporating the above implication, we apply our method to empirical data from Japanese organizations.

\section{Model formulation}
\label{sec:model_formulation}

\subsection{Asset prices and investment strategies}

Let $( \Omega, \mathcal{F}, \left\{ \mathcal{F}_t \right\}_{t\geq0}, \mathbb{P})$ be a filtered probability space with usual conditions.
We invest one risk-free asset and $m$ types of risky assets, and denote their prices at time $t$ by $S^0$ and $S_t \in \mathbb{R}^m$, respectively.
The price processes $S^0=\{S^0_t\}_{t\geq 0}$ and $S=\{S_t\}_{t\geq 0}$ are governed by the following:
\begin{align*}
    &\begin{cases}
        \displaystyle \frac{dS^0_t}{S^0_t} = r(t) dt,\\
        S^0_0 = s_0^0 >0,
    \end{cases}
	\hspace*{2em}
    &\begin{cases}
        \displaystyle \frac{dS^i_t}{S^i_t} = b^i(t) dt + \sum_{j=1}^m \sigma^{ij}(t) dW^j_t , \\
        S^i_0 = s^i_0 >0 , 
    \end{cases}
    \hspace*{1em} i=1,2,\cdots ,m,
\end{align*}
where 
$\left\{ W_t \right\}_{t \geq 0}$ is a $m$-dimensional standard Brownian motion,
$r: [0,T] \rightarrow \mathbb{R}^+$, $b: [0,T] \rightarrow \mathbb{R}^m$, and 
$\sigma: [0,T] \rightarrow \mathbb{R}^{m\times m}$ are deterministic continuous functions, and $T<\infty$ stands for the maturity.
We assume that all return rates of risky assets are larger than the risk-free rate, i.e.,
    \[
        b^i(t)-r(t)>0, \hspace*{2em} i \in \{1,2,\cdots,m\}, \; t \in [0,T].
    \]

We denote our investment strategy by $\{\pi_t\}_{0\leq t \leq T} \in \mathcal{A}^{\bar{\pi}}$.
A class of portfolio strategies $\mathcal{A}^{\bar{\pi}}$ is the collection of $\mathbb{R}^m$-valued $\mathcal{F}_t$-adapted process 
$\left\{ u_t \right\}_{0\leq t \leq T}$ which satisfies the following:
\begin{itemize}
    \item  $0 \leq u_t$,
    \item  $u_t ^\top \mathbbm{1} \leq \bar{\pi}$,
\end{itemize}
for any $t \in [0,T]$, where $\mathbbm{1} = (1,\cdots,1)^\top \in \mathbb{R}^m$.
The former condition prohibits short selling, while the latter limits the sum of weights to $\bar{\pi}$.
If $\bar{\pi} > 1$ leverage is allowed.

Then an investor's wealth $X_t$ satisfies the following:
\begin{align*}
    &\begin{cases}
        \displaystyle \frac{dX_t}{X_t} = \sum_{i=1}^m \pi^i_t  \frac{dS^i_t}{S^i_t} + \left( 1-  \pi_t^\top \mathbbm{1} \right) \frac{dS^0_t}{S^0_t},\\
        X_0 = x_0 = s^0_0 + s_0^\top \mathbbm{1}.
    \end{cases}
\end{align*}

\subsection{Performance criterion and value function}
As mentioned in the Introduction, we evaluate the strategies using a criterion based on the LMSE:
	\begin{equation}
		J^\pi_t(x) =  \mathbb{E} \left[  \frac{1}{2} (f(T)-X_T)_+^2 +  \left. \frac{1}{2} \frac{1}{T} \int_t^T (f(s)-X_s)_+^2 ds  \right| X_t = x\right].
		\label{eq:def_performance}
	\end{equation}
where $f: [0,T] \rightarrow \mathbb{R}$ is a deterministic function, and $(x)_+=x1_{\{x>0\}}$ for $x \in \mathbb{R}$.
The predetermined target wealth level is described by $f(t)$.
For instance, if $f(t)=(1-\bar{r}t)x_0$, the parameter $\bar{r}$ stands for the required absolute return of the investor.

We define the value function $V$ as follows:
	\[
		V_t(x) = \inf_{\pi \in \mathcal{A}^{\bar{\pi}}} J^\pi_t(x), \hspace*{1em}  t \in [0,T], \; x \in \mathbb{R}^+.
	\]
The value function satisfies the HJB equation (see e.g., \autocite{pham2009continuous})
	\begin{equation}
		\partial_t V_t(x) + \min_{\pi \in A^{\bar{\pi}}}\left\{ \mathscr{L}_t^{\pi}  V_t(x) \right\} = 0, \label{eq:HJB}
	\end{equation}
with the terminal condition
	\[
		V_T(x) = \frac{1}{2} (f(T)-x)_+^2,
	\]
where $\mathscr{L}_t^{\pi}$ is an infinitesimal generator of the process $X$
		\begin{align*}
			\mathscr{L}_t^{\pi} \phi(x) &=
				 \left( r(t) +  \left( b(t) -  r(t) \mathbf{1} \right)^\top \pi \right)x \partial_x \phi(x) \\
					& \hspace*{2em} + \frac{1}{2} x^2 \pi^\top \sigma(t)\sigma(t)^\top \pi \partial_{xx}\phi(x) + \frac{1}{2} \frac{1}{T} \left( f(t)-x \right)_+^2,
		\end{align*}
for $\phi : \mathbb{R} \rightarrow \mathbb{R}$ and 
\[
	A^{\bar{\pi}} = \left\{ \pi \in \mathbb{R}^m \;| \; \pi \geq 0, \; \pi^\top \mathbbm{1} \leq \bar{\pi}  \right\} .
\]
Because $\displaystyle \min_{\pi \in A^{\bar{\pi}}}\left\{ \mathscr{L}_t^{\pi}  V_t(x) \right\} $ is a quadratic programming problem about $\pi$, including the constraints, the analytical solution appears to be unavailable in class $A^{\bar{\pi}}$.

\subsection{Boundary conditions}
By the definition of the performance criterion, i.e., equation (\ref{eq:def_performance}),
 the boundary conditions of $V$ are given as follows:
\begin{align*}
	&V_t(0) = \frac{1}{2} \left( f(T) \right)_+^2 + \frac{1}{2} \frac{1}{T} \int_t^T \left( f(s) \right)_+^2 ds ; \\
	&V_t(x^*) = 0,\\
	&x^* = \max_{t \in [0,T]} f(t).
\end{align*}
If $X_t =0$, $X_s = 0$ for any $s \in (t, T]$ and $\pi \in \mathcal{A}^{\bar{\pi}}$.
Hence, we obtain the former condition.
Next, suppose that $X_t=x^*$.
Let us choose $\pi_s =0$ for any $s \in (t, T]$, and then $X_s = X_t e^{\int_t^s r(u)du}$.
Because $r$ is non-negative, $e^{\int_t^s r(u)du} > 1$.
This means that $X_s > x^* = \max_{t \in [0,T]} f(t)$.
Thus, we find $J^\pi_s(x)=0$, $x\geq x^*$ which leads $V_s(x)=0$, $x\geq x^*$.

\section{Numerical Method}
\label{sec:num_method}
As mentioned in the previous section, the analytical solution to the HJB equation (\ref{eq:HJB}) appears to be unavailable.
Hence, we solve it numerically using the kernel-based collocation method proposed by \autocite{Kansa1990}.
Precisely, we use the generalized version whose convergence is proved by \autocite{Nakano2014}.

\subsection{Interpolation function}
The kernel-based collocation method uses an interpolation function of scattered data.
Here, we briefly mention the interpolation using radial basis functions (RBFs).
Interested readers are referred to \autocite{wendland2004scattered} for details.

Let $(x^{(i)}, y_i)_{i=1,\cdots N}$, $x^{(i)}, y_i \in \mathbb{R}$ be the sample data,
 and $\mathcal{I}_{X,y}$ be the corresponding interpolation function, where $N$ denotes the number of samples.
The function $\mathcal{I}_{X,y}$ is defined by a linear combination of the kernel function $\Phi$:
        \[
            \mathcal{I}_{X,y}(x) = \sum_{i=1}^N \xi^{X,y}_i \Phi(x, x^{(i)})
        \]
where $\Phi: \mathbb{R}\times\mathbb{R} \rightarrow \mathbb{R},\; \xi^{X,y}_i \in \mathbb{R}$ for $i=1,\cdots,N$ are the interpolation weights.
We denote by $\xi^{X,y} = (\xi^{X,y}_1, \cdots, \xi^{X,y}_N)^\top$.
Then, the interpolation weights are determined by
        \[
        \xi^{X,y} = \left( R^{\Phi, X} \right)^{-1} y,
        \]
where $R^{\Phi, X} \in \mathbb{R}^{N \times N}$, $R^{\Phi, X}_{ij}=\Phi(x^{(i)}, x^{(j)})$.
As mentioned above, we choose a RBF $\phi$ as the kernel function:
    \[
        \Phi(x, x^{(i)}) = \phi(\| x-x^{(i)} \|)
    \]
Here, we employ the multiquadric function, a well-investigated RBF for interpolation (see, e.g., \autocite{Hardy1990}),
    \[
        \phi(r) = \sqrt{1+\frac{r^2}{\varepsilon^2}},
    \]
where $\varepsilon\in\mathbb{R}$ is called a shape parameter.

\subsection{Procedure for solving the HJB equation (\ref{eq:HJB})}

Here, we show the procedure for solving the HJB equation (\ref{eq:HJB}).
Let $\left\{ t_k \right\}_{k=0}^M$, $t_k = h_t k$, be the time grids,
 where $M \in \mathbb{N}$ is the number of time grids and $h_t=T/M$ is the time stepping size.
To use the left boundary condition, we take the sample points $x^{(i)} = (i-1) h_x$, where $h_x$ is the equal interval of the sample points.
    
    \begin{description}
        \item[Step1] Set $k=M$.
        \item[Step2] Set $\tilde{v}^k_i = \frac{1}{2} \left( f(T)-x^{(i)} \right)_+^2, \quad i=1,\cdots,N$.
        \item[Step3] Define interpolation function $\mathcal{I}_{X,\tilde{v}^k}$.
        \item[Step4] Calculate $\tilde{H}^k \in \mathbb{R}^{N}$ by
        \[
            \tilde{H}^k_i = \min_{\pi \in A^{\bar{\pi}}}\left\{ \mathscr{L}_t^{\pi}  \mathcal{I}_{X,\tilde{v}^k}\left( x^{(i)} \right) \right\}, \quad i= 1,\cdots, N. 
        \]
        \item[Step5] Define interpolation function $\mathcal{I}_{X,\tilde{H}^k}$.
        \item[Step6] Calculate $\tilde{v}^{k-1}_i \in \mathbb{R}^{N}$ by
        \[
            \tilde{v}^{k-1}_i = \mathcal{I}_{X,\tilde{v}^k}( x^{(i)} ) - h \mathcal{I}_{X,\tilde{H}^k}( x^{(i)} )
        \]
        \item[Step7] Update boundary values:
        \begin{align*}
            \tilde{v}^{k-1}_0 &= \frac{1}{2} \left( f(T) \right)_+^2 + \frac{1}{2} \frac{1}{T} \int_{t_{k-1}}^T \left( f(s) \right)_+^2 ds \\
            \tilde{v}^{k-1}_i &=  0, \quad \mathrm{if} \; x^{(i)} \geq x^*
        \end{align*}
        \item[Step8] Obtain the approximated solution $v^h$ at time step $k-1$ by
        \[
        v^h (t_{k-1}, \cdot) = \mathcal{I}_{X,\tilde{v}^{k-1}}(\cdot)
        \]
        \item[Step9] If $k=1$ finish, else set $k\leftarrow k-1$ and go back to Step4.
    \end{description}

\subsection{Remarks on implementation}
\label{sec:num_method_rmk}
Note that Step7 in the above procedure is different from the ordinary way in which the condition is given by $\tilde{v}^{k-1}_N = 0$ for $x^{(N)}=Nh_x=x^*$.
This change is the critical factor in stabilizing the numerical solution.
Let us recall the terminal condition and calculate its second derivative:
\[
    \partial_{xx} V_{T}(x) = \partial_{xx} \left( \frac{1}{2} (f(T)-x)_+^2 \right) = \mathbbm{1}_{ \{ x \leq x^* \} }.
\]
Hence, $\partial_{xx} V_{T}$ is discontinuous at $x=x^*$.
Because the approximated second derivative $\partial_{xx} v^h$ is a smooth function, it tends to behave unstably around the discontinuous point.
One technique to ease the instability is increasing the sample points.
Thus we add the extra sample points outside the right boundary $x^*$; that is, $N$ is determined to satisfy $Nh_x> x^*$.
Because $V_t(x)$ is defined on $x\in\mathbb{R}^+$ and $V_t(x)=0$, $x\geq x^*$, this expansion of the computational domain is valid.
In section \ref{sec:check_vf}, we observe the unstable result with the ordinal method and the stabilized result with our method.

Our program is written in Python except for Step4, the highest computational load step.
Step4 is the optimization of each sample point, and they are able to proceed independently.
Hence, we implement Step4 in C++ to solve the quadratic programming parallelly.
The quadratic programming is solved by OSQP \autocite{osqp}.
The numerical solution in the following section is obtained in a few minutes.\footnote{The computer we use has the following specifications: Intel Core i9-X10900X, 3.5Hz, 12Cores, and 32GB RAM.}

\section{Numerical Results}
\label{sec:num_results}

In this section, we show two numerical results with:
(i)  artificial data to investigate the general properties of the optimal portfolio obtained by our numerical method; and
(ii) empirical data used in \autocite{Ieda2013a} and \autocite{Ieda2014} to obtain the optimal portfolio under the leverage constraint, which has not been investigated well in the literature.

We evaluate our numerical optimal portfolio using Monte-Carlo simulations with $10^5$ sample paths.
We mainly observe following statistics:
\begin{description}
	\item[Mean wealth] $ \displaystyle \bar{X}_t := \frac{1}{N^M} \sum_{i=1}^{N^M} X^{(i)}_t$
	\item[Achievement rate] $\displaystyle A_t := \frac{1}{N^M} \sum_{i=1}^{N^M} \mathbbm{1}_{ \left\{ X^{(i)}_t \geq f(t) \right\} }$
	\item[95-percentile point] $P_t$
	\item[Tracking error rate\footnote{We see the histogram of $E^{(i)}_t$.}] $E^{(i)}_t := \dfrac{X^{(i)}_t - f(t)}{f(t)}$ 
\end{description}
where $X^{(i)}_t$ is the $i$-th sample path, and $N^M=10^5$ stands for the number of simulation paths.
In the simulations, we allow for the monthly rebalance of the portfolio weight without any mention.

\subsection{Artificial data}

\subsubsection*{Calculation setting}

The invested assets consist of a risk-free asset, a low-risk asset $S^1$ and a high-risk asset $S^2$ with the following parameters:
\begin{align*}
	& r= 0.01\%,\\
	&b(t) = (1\%, 5\%)^\top,\\
	&\sigma(t)\sigma(t)^\top = 
		\left( \begin{array}{cc}
			\sigma_1	&  0 \\
			0 & \sigma_2
		\end{array}	\right)
		\left( \begin{array}{cc}
			1	&  \rho \\
			\rho & 1
		\end{array}	\right)
		\left( \begin{array}{cc}
			\sigma_1	&  0 \\
			0 & \sigma_2
		\end{array}	\right),\\
	&\sigma_1 = 1\%, \; \sigma_2 = 20\%, \rho=-0.3.
\end{align*}
The target wealth level is as follows:\footnote{Thus, the boundary point is given by $x^* = (1+\bar{r}T)x_0$.}
\[
	f(t) = (1 + \bar{r} t) x_0,
\]
that is, we require the absolute return $\bar{r}$, as mentioned in Introduction.
We set the initial wealth as $x_0=100$ and terminal time as $T=10$ [years].

The parameters to obtain the numerical solution are as follows:
\begin{table}[H]
	\centering
	\caption{Parameters for numerical solutions (artificial data)}
	\begin{tabular}{ccl}
		\toprule
		Symbol & Value & Description \\
		\midrule
		$M$ &  6$\times10^4$  & terminal time $\times$ monthly $\times$ 500 \\
		$h_x$ &  0.5  & - \\
		$N$ &  $\dfrac{x^*}{h_x} + 6$  & locate five extra points outside $x^*$  \\
		$\varepsilon$ &  $0.5h_x$  & shape parameter of RBF \\
		\bottomrule
	\end{tabular}
\end{table}

\subsubsection*{Value function and portfolio weight}
\label{sec:check_vf}

\begin{figure}[H]
	\begin{minipage}[t]{0.33\columnwidth}
		\centering
		\includegraphics[width=12em]{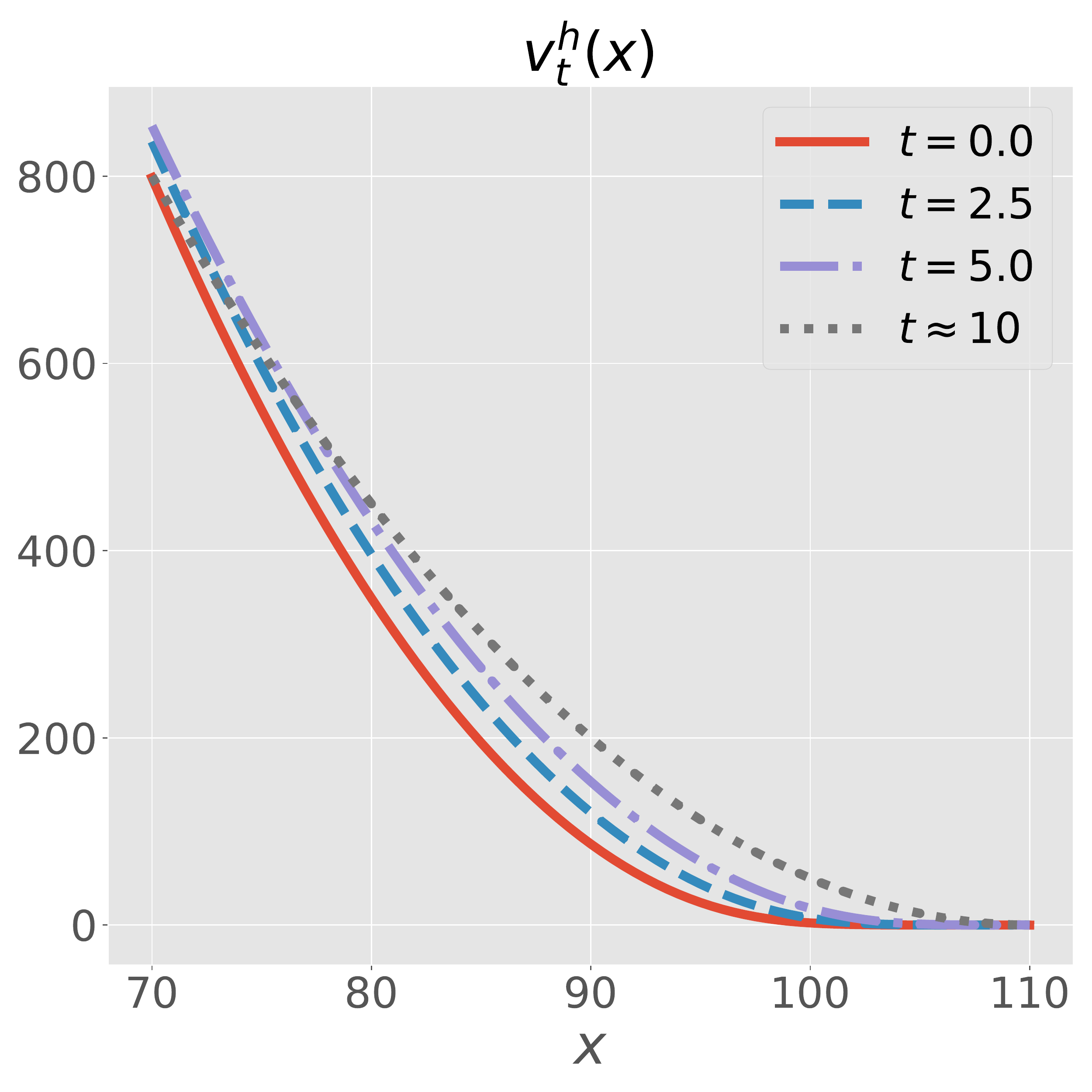}
	\end{minipage}
	\begin{minipage}[t]{0.33\columnwidth}
		\centering
		\includegraphics[width=12em]{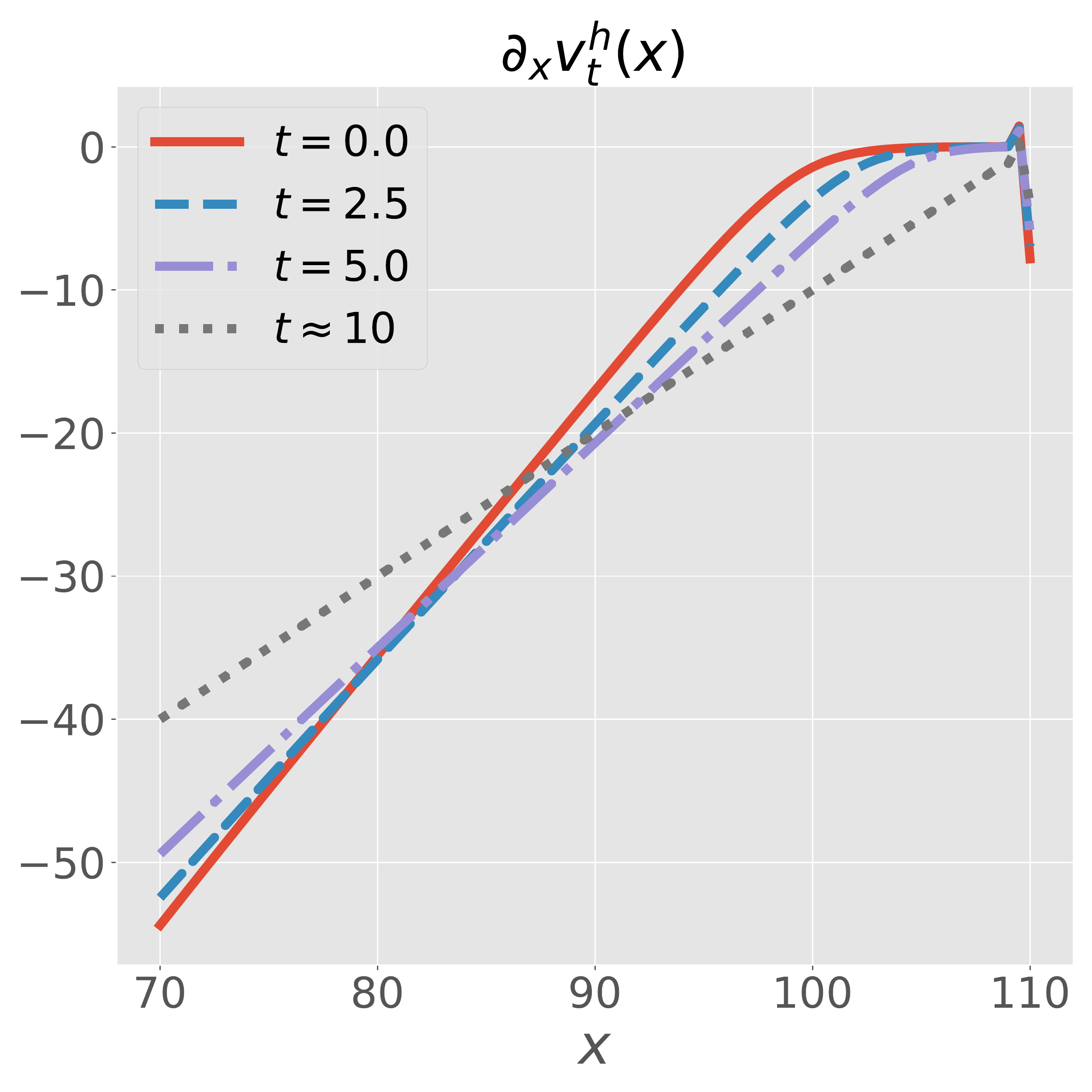}
	\end{minipage}
	\begin{minipage}[t]{0.33\columnwidth}
		\centering
		\includegraphics[width=12em]{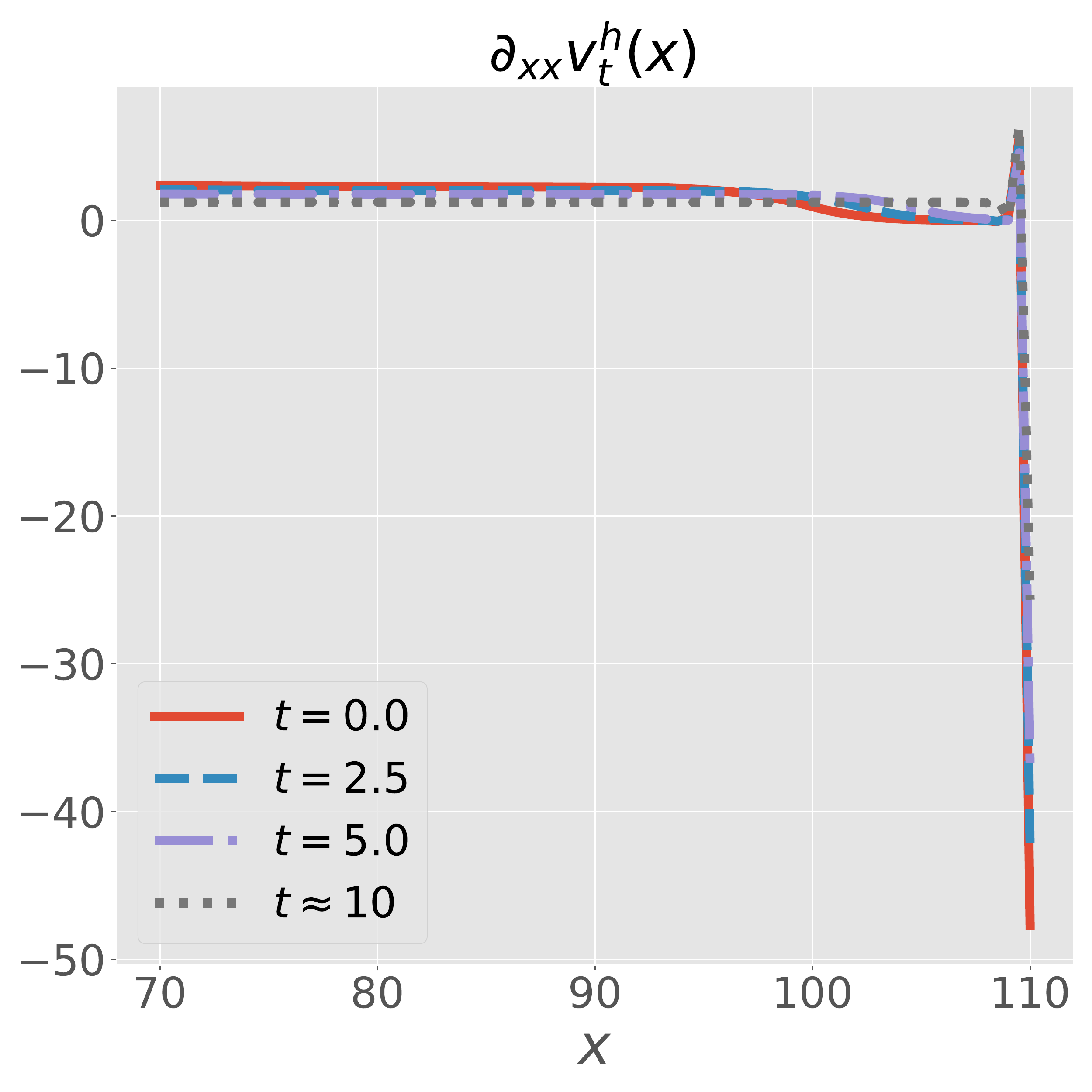}
	\end{minipage}
	\caption{ Numerical solutions of the value function and its derivatives
				 with parameters $\bar{r}=1\%$, $\bar{\pi}=1$ and $N=\frac{x^*}{h_x}+1$}
	\label{fig:vf_ng}
\end{figure}

\begin{figure}[H]
	\begin{minipage}[t]{0.33\columnwidth}
		\centering
		\includegraphics[width=12em]{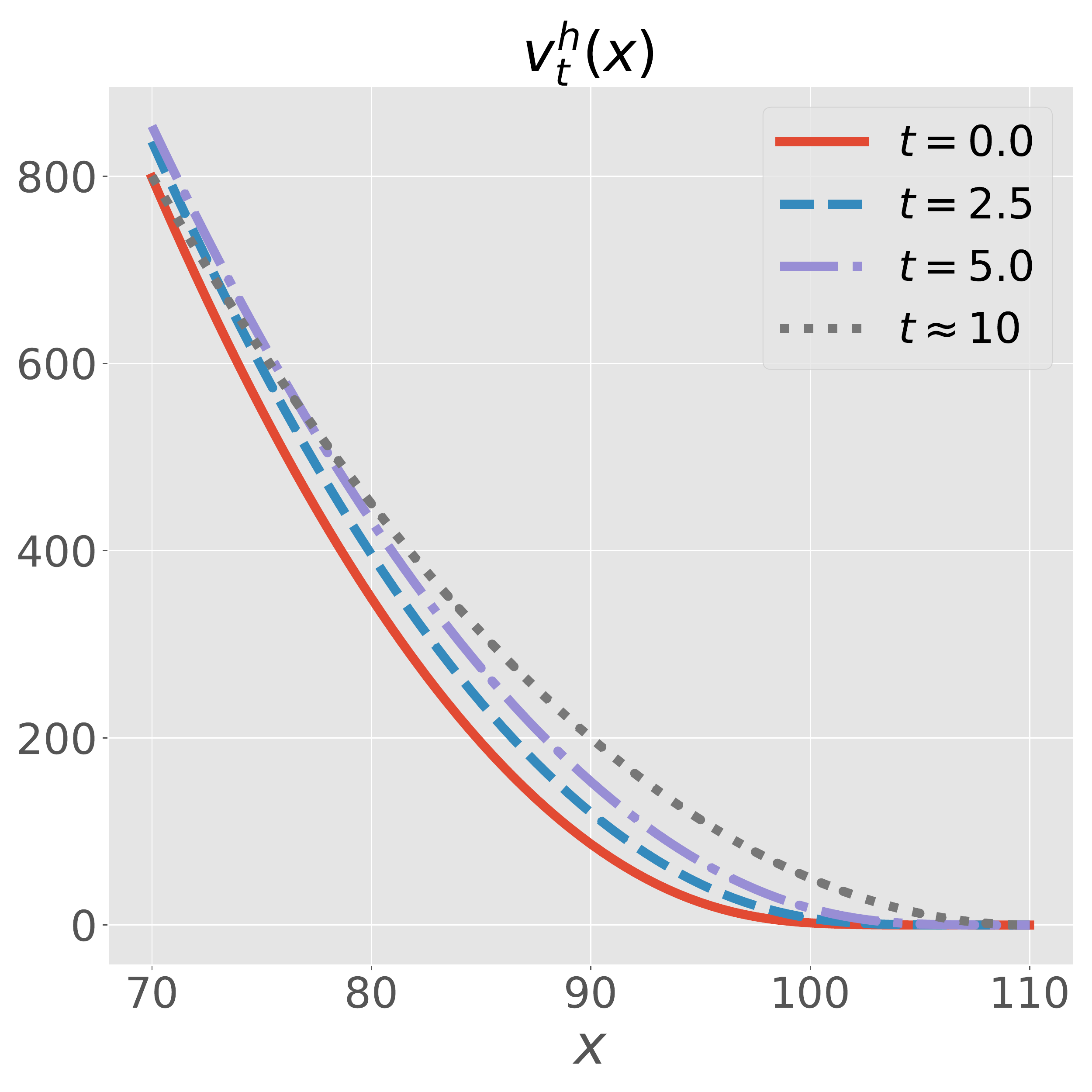}
	\end{minipage}
	\begin{minipage}[t]{0.33\columnwidth}
		\centering
		\includegraphics[width=12em]{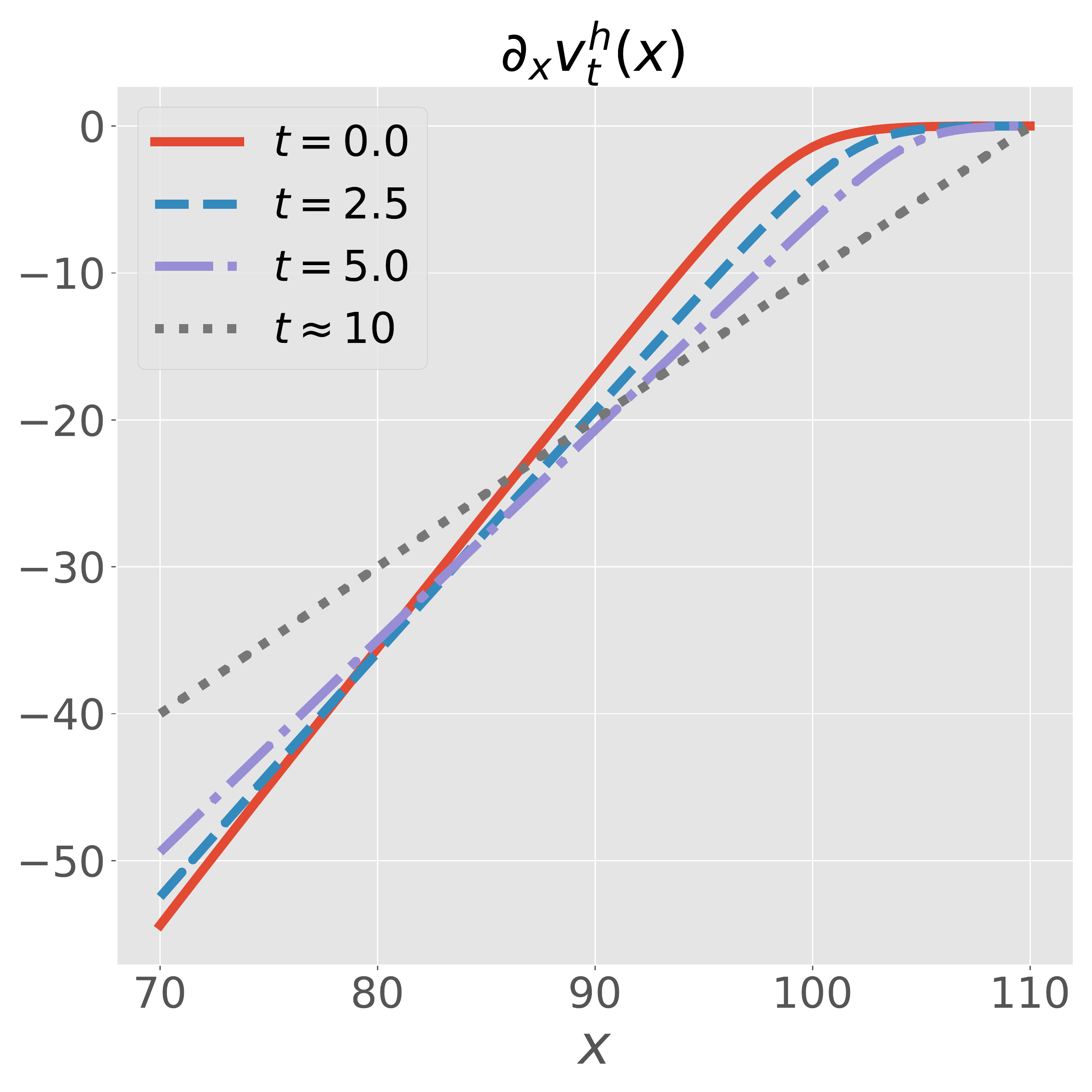}
	\end{minipage}
	\begin{minipage}[t]{0.33\columnwidth}
		\centering
		\includegraphics[width=12em]{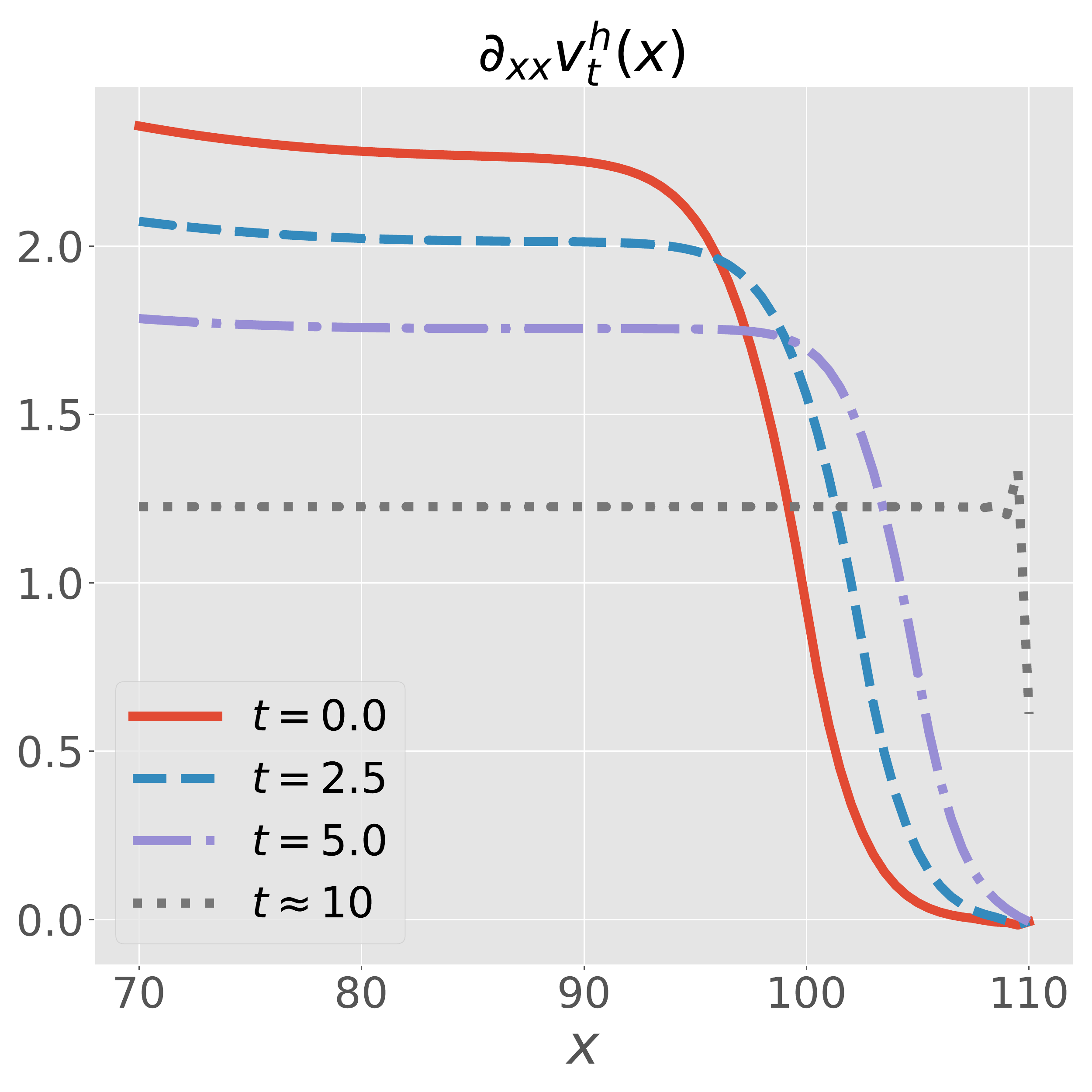}
	\end{minipage}
	\caption{ Numerical solutions of the value function and its derivatives with parameters $\bar{r}=1\%$ and $\bar{\pi}=1$. }
	\label{fig:vf}
\end{figure}

We first check whether the numerical solution of the value function is obtained with sufficient accuracy.
Figure\ref{fig:vf_ng} shows the obtained value function and its derivatives with parameters $\bar{r}=1\%$, $\bar{\pi}=1$, and $N=\frac{x^*}{h_x}+1$.
Obviously, the approximated derivatives behave unstably around the boundary point $x^*$.
Because the strategy $\pi$ highly depends on the sign of the second derivative,
 we cannot obtain a reliable strategy with the ordinary boundary condition.
Figure \ref{fig:vf} shows the same solutions with parameter $N=\frac{x^*}{h_x}+6$.
As mentioned in Section \ref{sec:num_method_rmk}, we cope with the discontinuity around point $x^*$ by locating the extra sample points.
We find that the approximated derivatives are stabilized.
Moreover, the results in the following subsections show that the approximated value function and its derivatives appear to be obtained with sufficient accuracy.

\begin{figure}[H]
	\centering
	\includegraphics[width=34.5em]{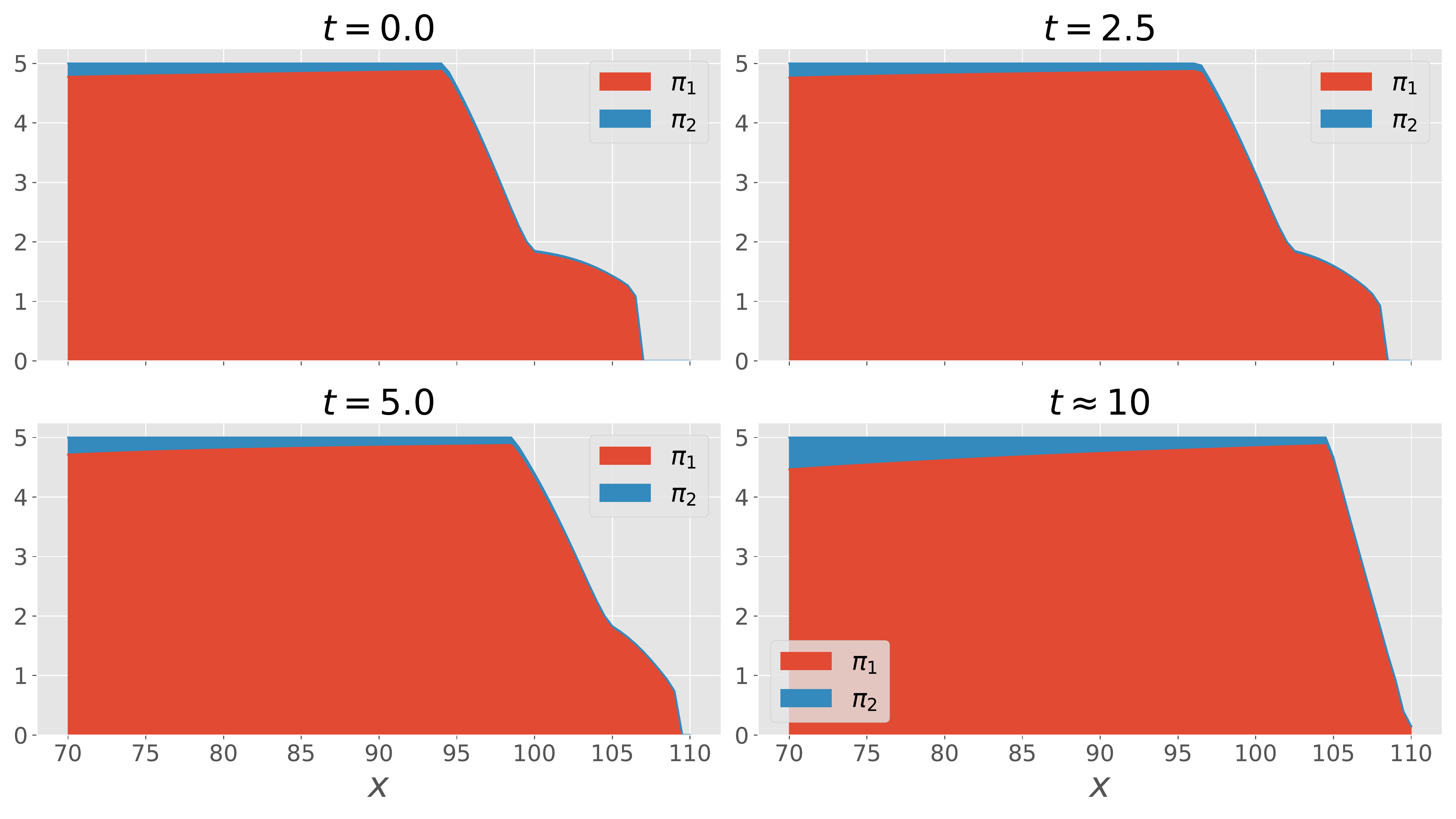}
	\caption{ The optimal portfolio weight with parameters $\bar{r}=1\%$ and $\bar{\pi}=5$. }
	\label{fig:art_stra_pi_5}
\end{figure}

\begin{figure}[H]
	\centering
	\includegraphics[width=34.5em]{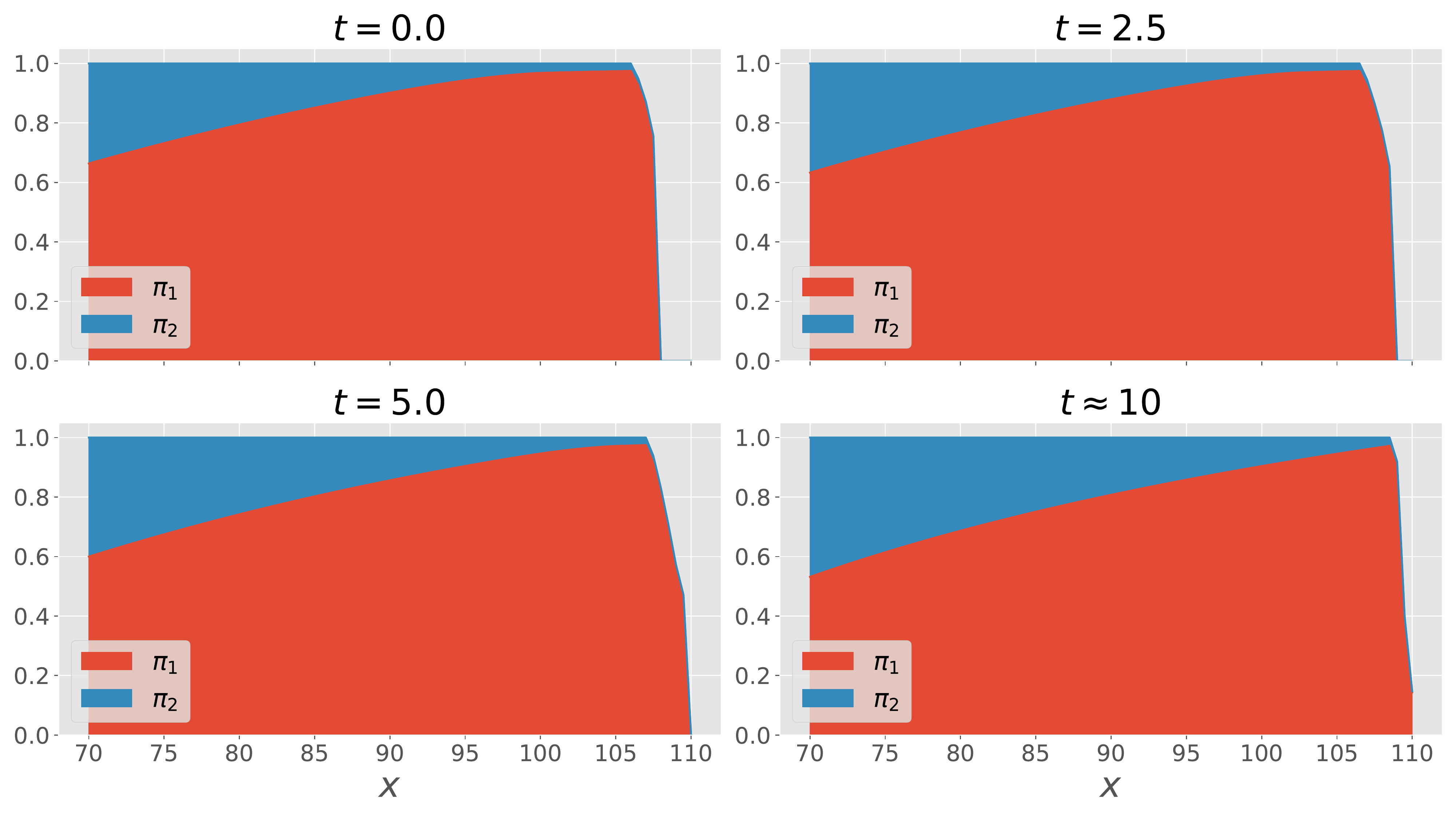}
	\caption{ The optimal portfolio weight with parameters $\bar{r}=1\%$ and $\bar{\pi}=1$. }
	\label{fig:art_stra_pi_1}
\end{figure}

The optimal portfolio weight obtained by our numerical method is shown in Figures \ref{fig:art_stra_pi_5} and \ref{fig:art_stra_pi_1}.
We see that the optimal weight is obtained stably.
The difference between Figures \ref{fig:art_stra_pi_5} and \ref{fig:art_stra_pi_1} is the value of the allowed leverage.
Figure \ref{fig:art_stra_pi_5} indicates that the main factor determining the portfolio weight is leverage, and the balance between the assets $S^1$ and $S^2$ is the secondary contribution.
If leverage is allowed, the required profit amount, not the profit rate, can be covered by the large exposure of the low-risk asset funded by leverage.
Hence, there is no incentive to invest in the high-risk asset to earn profit.
The high-risk asset is included to reduce the portfolio risk through the correlation with the low-risk asset.
In contrast,  if leverage is restricted, there is no choice to increase the weight of the high-risk asset to obtain the desired profit, as shown in Figure \ref{fig:art_stra_pi_1}.

\subsubsection*{Simulation results}

\begin{figure}[H]
	\centering
	\includegraphics[width=35em]{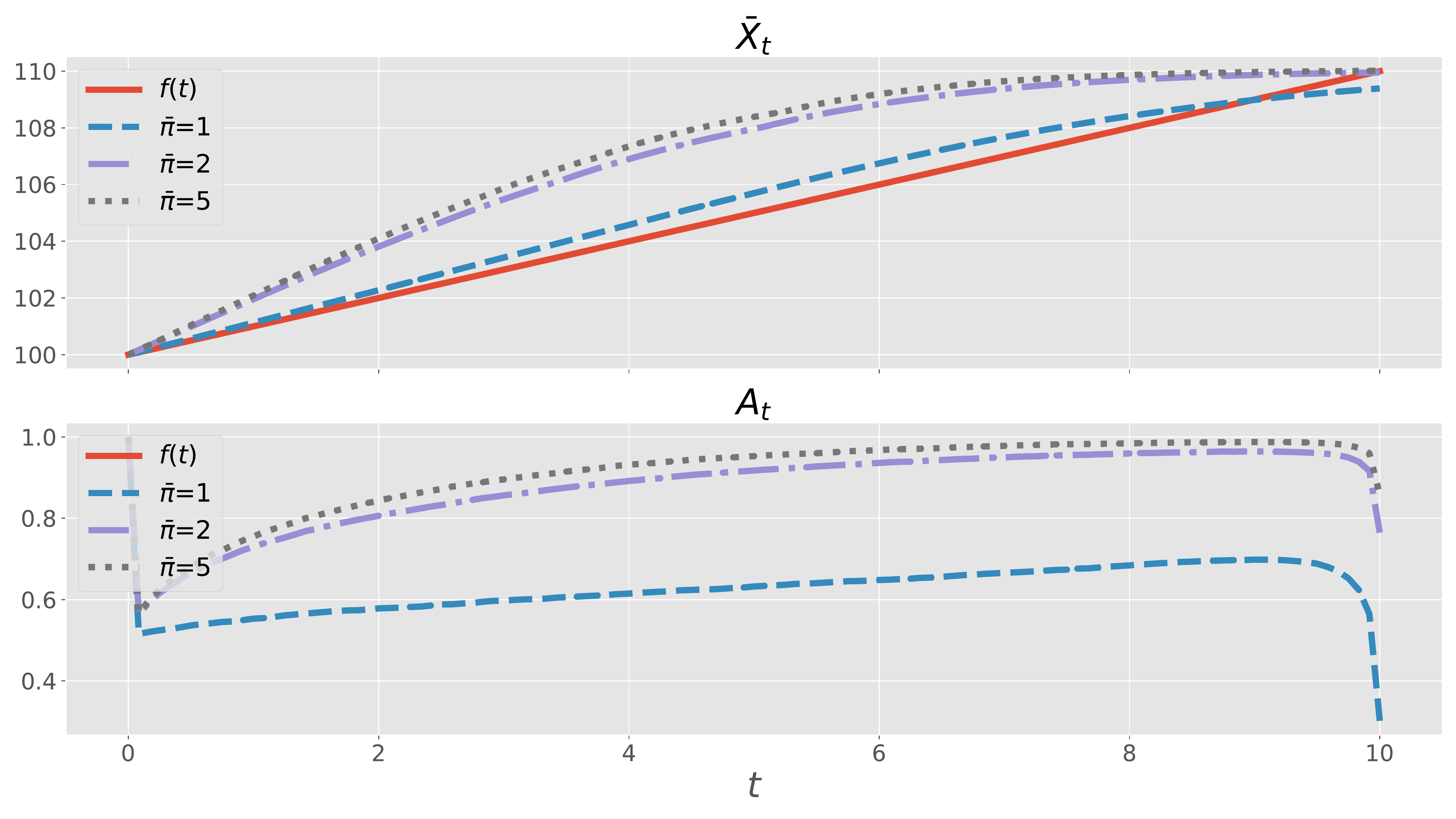}
	\caption{ Simulation results $\bar{X}_t$ and $A_t$ with parameters $\bar{r}=1\%$. }
	\label{fig:art_sim_res_wo_margin}
\end{figure}

Figure \ref{fig:art_sim_res_wo_margin} shows the mean wealth $\bar{X}_t$ and the achievement rate $A_t$ of the simulation with parameter $\bar{r}$=1\%.
The mean wealth indicated in the upper panel appears to be superior to the target wealth regardless of the leverage constraint.
However, the lower panel shows that the achievement rate plummets close to the terminal time.

\begin{figure}[H]
	\centering
	\includegraphics[width=34em]{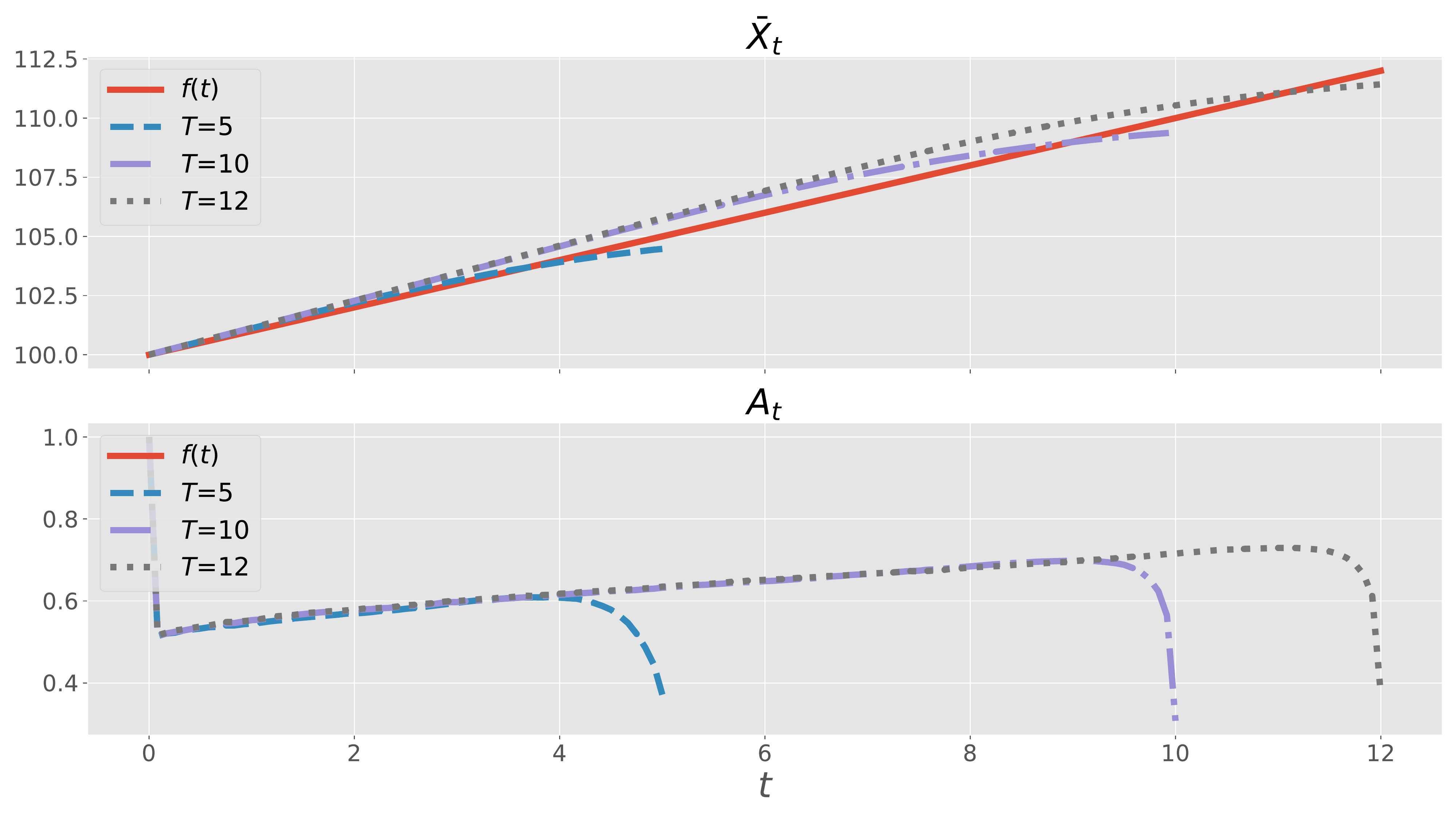}
	\caption{  Simulation results $\bar{X}_t$ and $A_t$ with parameters $\bar{r}=1\%$ and several terminal times. }
	\label{fig:art_sim_res_comp_terminal}
\end{figure}

To figure out the details, we run simulations with different terminal times.
The results are shown in Figure \ref{fig:art_sim_res_comp_terminal}.
First, we see that the plummeting of achievement rate occurs independently of the terminal times.
Comparing the purple chained and the gray dotted lines in the lower panel of Figure \ref{fig:art_sim_res_comp_terminal},
 we find the more notable fact that the plummet at $t=10$ is avoided when we take $T=12$.
The plummet at $t=5$ is also avoided when $T=10$ or $T=12$.
This indicates that the plummet that occurs close to the terminal time disappears by taking a longer terminal time.

\begin{figure}[H]
	\centering
	\includegraphics[width=34em]{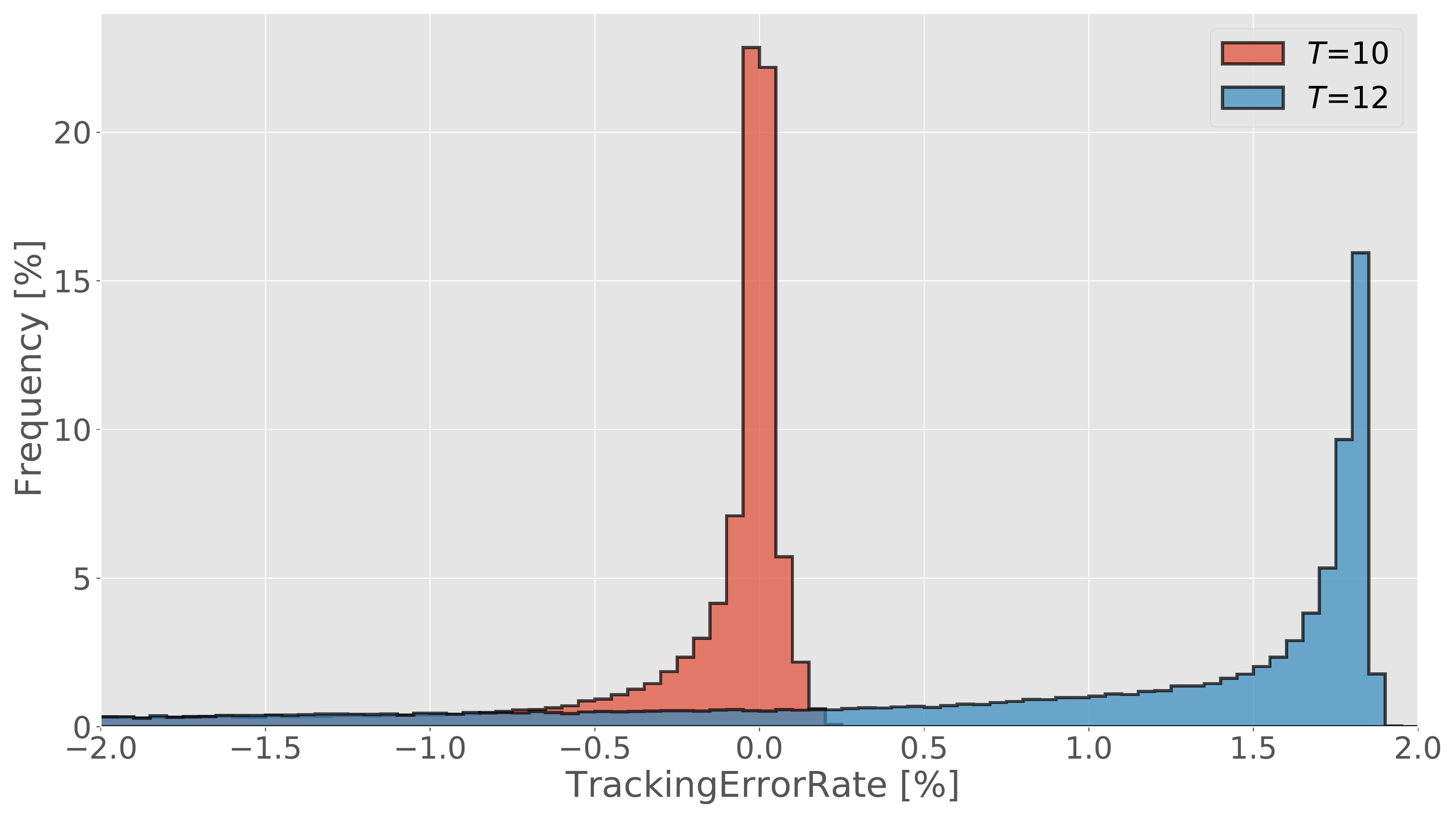}
	\caption{ Histogram of the tracking error rates at time $t=10$.
	}
	\label{fig:art_hist_comp_terminal}
\end{figure}

Hence the reason for the plummet is not the invested assets.
We can intuitively understand it through a feature of the LMSE.
Under the LMSE criterion, there is no incentive for investor wealth to exceed the required level.
However, the downside risk is not eliminated ideally, and thus, quite a few sample paths do not reach $f(T)$.
The histogram of the tracking error rate at $t=10$ displayed in Figure \ref{fig:art_hist_comp_terminal} supports this explanation,
 as indicated by the red area in the case of $T=10$.
Observing the blue area describing the case of $T=12$, we find that the peak of the histogram shifts to the right side.
This appears to be caused by the incentive exceeding $f(10)$ because investor wealth finally must reach $f(12) > f(10)$ in the case of $T=12$.

We state a more direct approach to control the incentive for exceeding the required wealth level by introducing the margin rate $\varepsilon^M$:
\begin{equation}
	f(t) = \left( 1+(\bar{r}+\varepsilon^M) t \right) x_0.
	\label{eq:target_wealth_with_margin}
\end{equation}
Table \ref{tb:art_comp_margin} shows the comparison of the statistics with various margin rates.
We observe the achievement rate $A_T$ and find that the plummets at the terminal time are avoided.
Hence, we use equation (\ref{eq:target_wealth_with_margin}) to calculate the value function and the portfolio weight in the rest of this subsection.
\begin{table}[H]
	\centering
	\caption{
		Comparison of the statistics with various margin rates.\\
		The target return is $\bar{r}=$1\% and the leverage rate is $\bar{\pi}$=1.
	}
	\label{tb:art_comp_margin}
	\begin{tabular}{llllll}
		\toprule
		$\varepsilon^M$ & $(1+\bar{r}T)x_0$ & $\bar{X}_T$ & $A_{T/2}$ & $A_T$ &  $P_T$ \\
		\midrule
		0.0\% &        &       109.4 &      63\% &  31\% &  106.5 \\
		0.1\% &        &       110.2 &      64\% &  73\% &  106.6 \\
		0.2\% &  110.0 &       110.7 &      65\% &  74\% &  106.6 \\
		0.3\% &        &       111.2 &      66\% &  75\% &  106.6 \\
		0.4\% &        &       111.6 &      67\% &  77\% &  106.5 \\
		0.5\% &        &       112.0 &      68\% &  78\% &  106.4 \\
		\bottomrule
	\end{tabular}
\end{table}

\subsubsection*{Simulation results with margin rate $\varepsilon^M$}
\begin{figure}[H]
	\centering
	\includegraphics[width=34em]{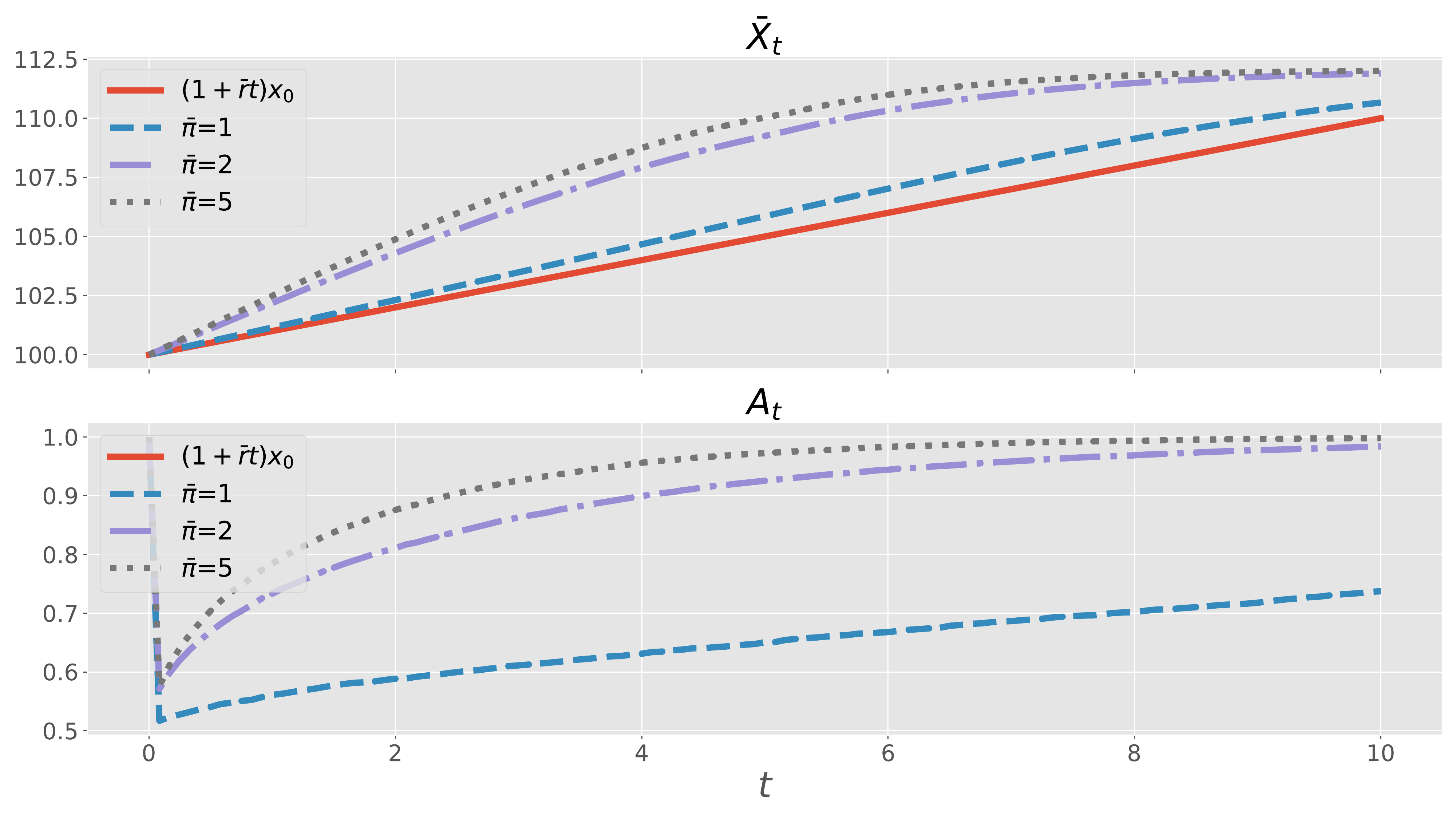}
	\caption{
		Simulation results $\bar{X}_t$ and $A_t$ with parameters $\bar{r}=1\%$ and $\varepsilon^M=0.2\%$.
	}
	\label{fig:art_sim_result_margin}
\end{figure}

Figure \ref{fig:art_sim_result_margin} shows the simulation results $\bar{X}_t$ and $A_t$ with parameters $\bar{r}=1\%$ and $\varepsilon^M=0.2\%$.
Compared to Figure \ref{fig:art_sim_res_comp_terminal}, we find that the achievement rates are improved.
Moreover, the achievement rate increases monotonically.
This fact implies that the long-term investment offers a stable return.

Table \ref{tb:art_comp_stat_sim} indicates the statistics with various target returns and leverage levels fixing the margin rate as $\varepsilon^M$ = 0.2\%.
If the target return is $\bar{r}=2\%$, the target wealth cannot be achieved without leverage.
One criterion to determine whether the optimal portfolio can earn the target return is to check the following two points: (i) the achievement rate increases, that is, $A_{T/2} < A_T$; (ii) the 95-percentile point $P_T$ does not diverge from the terminal wealth. 
 \begin{table}[H]
	\centering
	\caption{
		Comparison of the statistics with various target returns and leverage levels.\\
		The margin rate is $\varepsilon^M$ = 0.2\%.
	}
	\label{tb:art_comp_stat_sim}
	\begin{tabular}{llllll}
		\toprule
		$\bar{r}$ & $\bar{\pi}$ & $\bar{X}_T$ & $A_{T/2}$ &  $A_T$ &  $P_T$ \\
		\midrule
		&           1 &       110.7 &      65\% &   74\% &  106.6 \\
		1\% &           2 &       111.9 &      93\% &   98\% &  111.6 \\
		&           5 &       112.0 &      97\% &  100\% &  111.9 \\
		\midrule
		&           1 &       114.0 &      17\% &    7\% &  104.8 \\
		2\% &           2 &       120.4 &      65\% &   76\% &  113.3 \\
		&           5 &       122.3 &      94\% &   99\% &  121.6 \\
		\bottomrule
	\end{tabular}
	\hspace*{2em}
	\begin{tabular}{llllll}
		\toprule
		$\bar{r}$ & $\bar{\pi}$ & $\bar{X}_T$ & $A_{T/2}$ &  $A_T$ &  $P_T$ \\
		\midrule
		&           1 &       116.4 &       7\% &    0\% &  101.3 \\
		3\% &           2 &       125.3 &      32\% &   32\% &  113.0 \\
		&           5 &       132.1 &      89\% &   96\% &  130.7 \\
		\bottomrule
		\vspace*{3em}
	\end{tabular}

\end{table}
Following the above criteria, the optimal portfolio weight with $\bar{r}=2\%$ and $\bar{\pi}=2$ is a well-tracking strategy.
Meanwhile, it is difficult to earn the required return stably for the optimal weight with $\bar{r}=3\%$ and $\bar{\pi}=2$.
For the cases tested above, the optimal strategy with leverage $\bar{\pi}=5\%$ generates the target wealth stably.

\subsubsection*{Rebalance frequency}
\begin{figure}[H]
	\begin{minipage}[t]{0.48\columnwidth}
		\centering
		\includegraphics[width=18em]{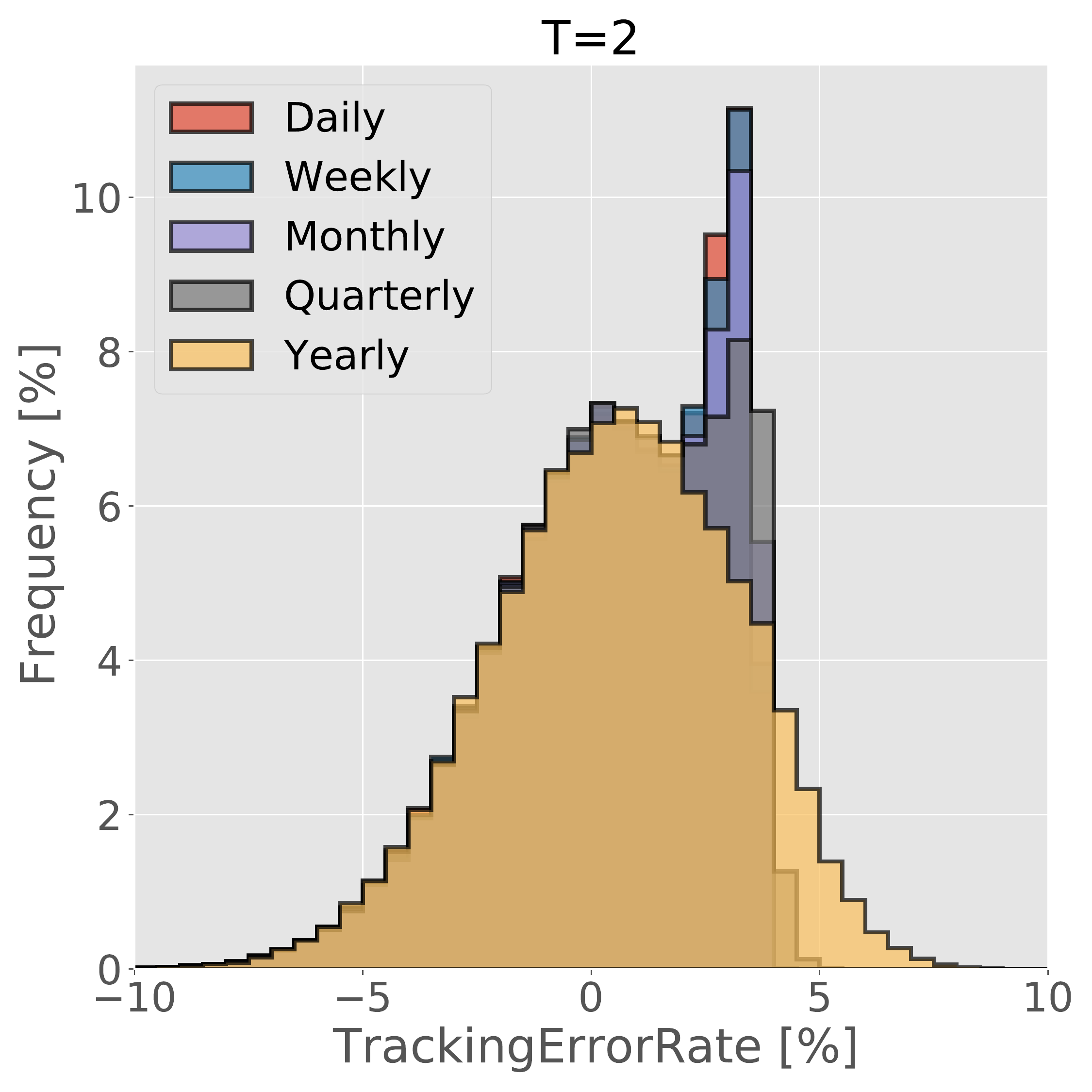}
	\end{minipage}
	\begin{minipage}[t]{0.48\columnwidth}
		\centering
		\includegraphics[width=18em]{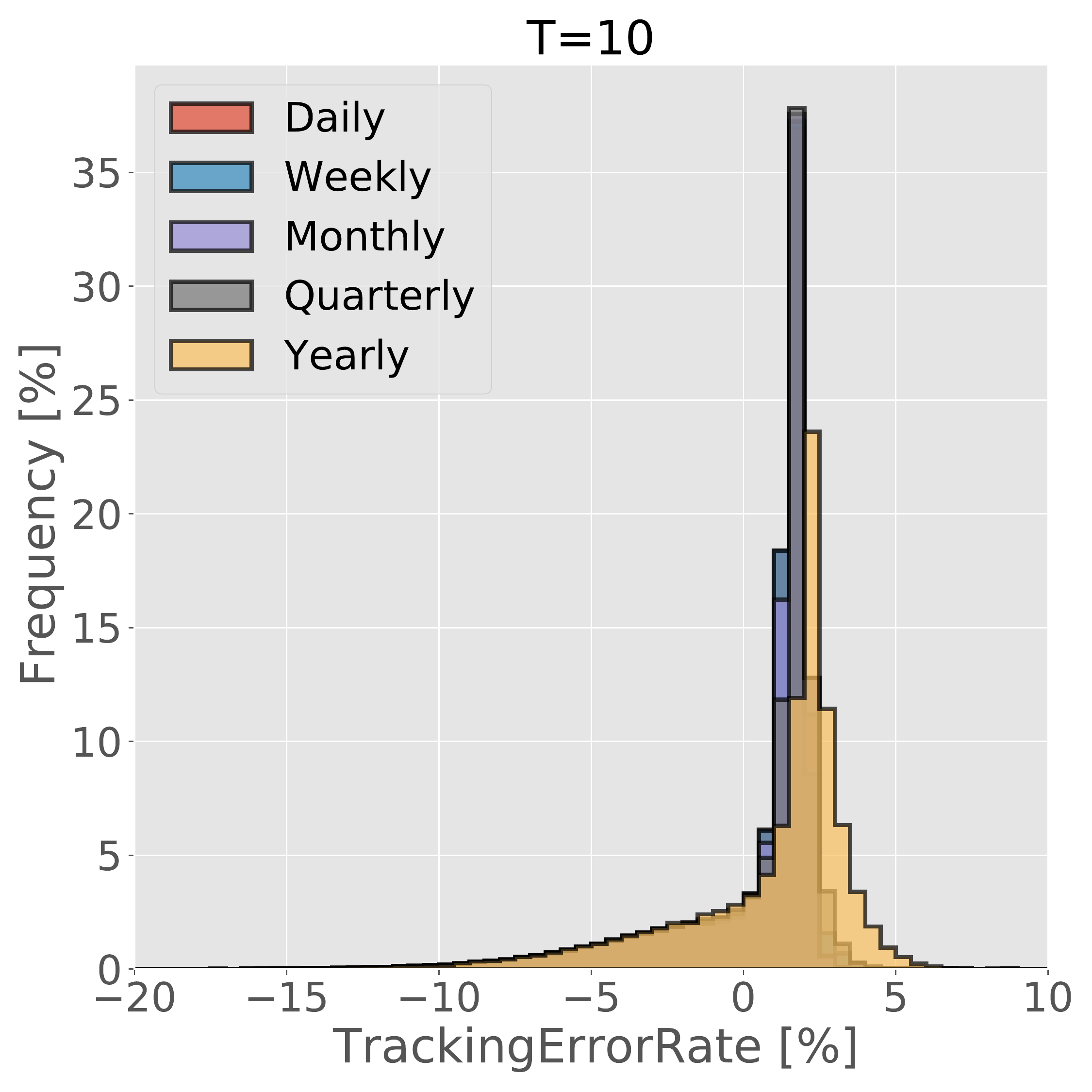}
	\end{minipage}
	
	\caption{ Histograms of the tracking error rates at the terminal time.}
	\label{fig:art_comp_freq}
\end{figure}

We check how the rebalance frequency affects investment performance.
Table \ref{tb:art_comp_freq} displays the comparison of the statistics with various rebalance frequencies.
This implies that there is no critical effect, even though we choose the yearly frequency.
However, Figure \ref{fig:art_comp_freq}, which describes the histogram of the error rate, shows that the result of the yearly rebalance, especially in the case of $T=2$, spreads broadly.
Thus, the accuracy of the control decreases certainly.
\begin{table}[H]
	\centering
	\caption{
		Comparison of the statistics with various rebalance frequencies.
		The target return is $\bar{r}=$2\% and the leverage rate is $\bar{\pi}$=2.
	}
	\label{tb:art_comp_freq}
\begin{tabular}{llllllll}
	\toprule
	$T$ & $\varepsilon^M$ &  rebalance & $(1+\bar{r}T)x_0$ & $\bar{X}_T$ & $A_{T/2}$ & $A_T$ &  $P_T$ \\
	\midrule
	&                 &      Daily &                   &       104.4 &      55\% &  59\% &   99.8 \\
	&                 &     Weekly &                   &       104.4 &      55\% &  59\% &   99.8 \\
	2 &              2\% &    Monthly &             104.0 &       104.4 &      56\% &  59\% &   99.9 \\
	&                 &  Quarterly &                   &       104.4 &      55\% &  59\% &   99.9 \\
	&                 &     Yearly &                   &       104.5 &      56\% &  59\% &   99.8 \\
	\midrule
	&                 &      Daily &                   &       120.4 &      65\% &  77\% &  113.3 \\
	&                 &     Weekly &                   &       120.4 &      65\% &  77\% &  113.3 \\
	10 &            0.2\% &    Monthly &             120.0 &       120.4 &      65\% &  76\% &  113.3 \\
	&                 &  Quarterly &                   &       120.5 &      65\% &  76\% &  113.3 \\
	&                 &     Yearly &                   &       121.0 &      64\% &  75\% &  113.2 \\
	\bottomrule
\end{tabular}
\end{table}

\subsection{Empirical data}

\subsection*{Calculation settings}
The investment assets consist of
 a risk-free asset $S^0$,
 an index of Japanese bonds $S^1$,
 that of Japanese stocks $S^2$,
 that of foreign bonds $S^3$,
 and that of foreign stocks $S^4$.
The parameters of the above assets are as follows:
\begin{align*}
	&r = 0.001\%\\
	&b(t) = (3\%, 4.8\%, 3.5\%, 5.0\%)^\top \\
	&\sigma(t)\sigma(t)^\top = 
		\begin{pmatrix}
			29.7&			18.2&		-4.39&		-5.41\\
			18.2&		495&			-77.8&		119\\
			-4.39&		-77.8&			181&		147\\
			-5.41&		119&		147&		394\\
		\end{pmatrix}
		\times 10^{-4}.
\end{align*}
We set the terminal time as $T=15$.
The target wealth is the difference between the estimated income $C(t)$ and the expense $B(t)$ of a Japanese pension fund.
Table \ref{tb:emp_bene_cont} shows the estimations of $B(t)$ and $C(t)$.
Then, we see that $x_0 = B(0)-C(0)=5.292$.
See \autocite{Ieda2013a} or \autocite{Ieda2014} for further information.
Note that $B(t)-C(t)$ is almost a linear function(see Figure \ref{fig:emp_sample_path} or Figure \ref{fig:emp_sim_res} displayed in the latter).
Approximately 7\% absolute return is required in this situation.
Based on the suggestion of the previous subsection, we determine the criterion $f(t)$ with the margin:
\[
	f(t) = 1.1 (B(t)-C(t)).
\]
The parameters to obtain the numerical solution are as in Table \ref{tb:emp_num_params}.

\begin{table}[H]
	\centering
	\caption{
		Estimated income and expense of the Japanese pension fund (billion yen)
	}
	\label{tb:emp_bene_cont}
	\begin{tabular}{cccc}
		\toprule
		$t$ & $B(t)$ & $C(t)$ & fiscal year \\
		\midrule
		0 & 67.286 & 61.993 & 2040 \\
		1 & 69.053 & 62.808 & 2041 \\
		2 & 70.718 & 63.611 & 2042 \\
		3 & 72.329 & 64.405 & 2043 \\
		4 & 73.887 & 65.193 & 2044 \\
		5 & 75.374 & 65.978 & 2045 \\
		6 & 76.803 & 66.766 & 2046 \\
		7 & 78.246 & 67.569 & 2047 \\
		\bottomrule
	\end{tabular}
	\hspace*{2em}
	\begin{tabular}{cccc}
		\toprule
		$t$ & $B(t)$ & $C(t)$ & fiscal year \\
		\midrule
		8 & 79.761 & 68.393 & 2048 \\
		9 & 81.331 & 69.241 & 2049 \\
		10 & 82.893 & 70.112 & 2050 \\
		11 & 84.437 & 70.996 & 2051 \\
		12 & 85.940 & 71.878 & 2052 \\
		13 & 87.432 & 72.759 & 2053 \\
		14 & 88.909 & 73.632 & 2054 \\
		15 & 90.322 & 74.490 & 2055 \\
		\bottomrule
	\end{tabular}
\end{table}
\begin{table}[H]
	\centering
	\caption{Parameters for numerical solutions (empirical data)}
	\begin{tabular}{ccl}
		\toprule
		Symbol & Value & Description \\
		\midrule
		$M$ &  9,000  & terminal time $\times$ monthly $\times$ 50 \\
		$h_x$ &  0.5  & - \\
		$N$ &  $\dfrac{x^*}{h_x} + 6$  & locate five extra points outside $x^*$  \\
		$\varepsilon$ &  $0.5h_x$  & shape parameter of RBF \\
		\bottomrule
	\end{tabular}
	\label{tb:emp_num_params}
\end{table}

\subsection*{Optimal portfolio weights}
\begin{figure}[H]
	\centering
	\includegraphics[width=35em]{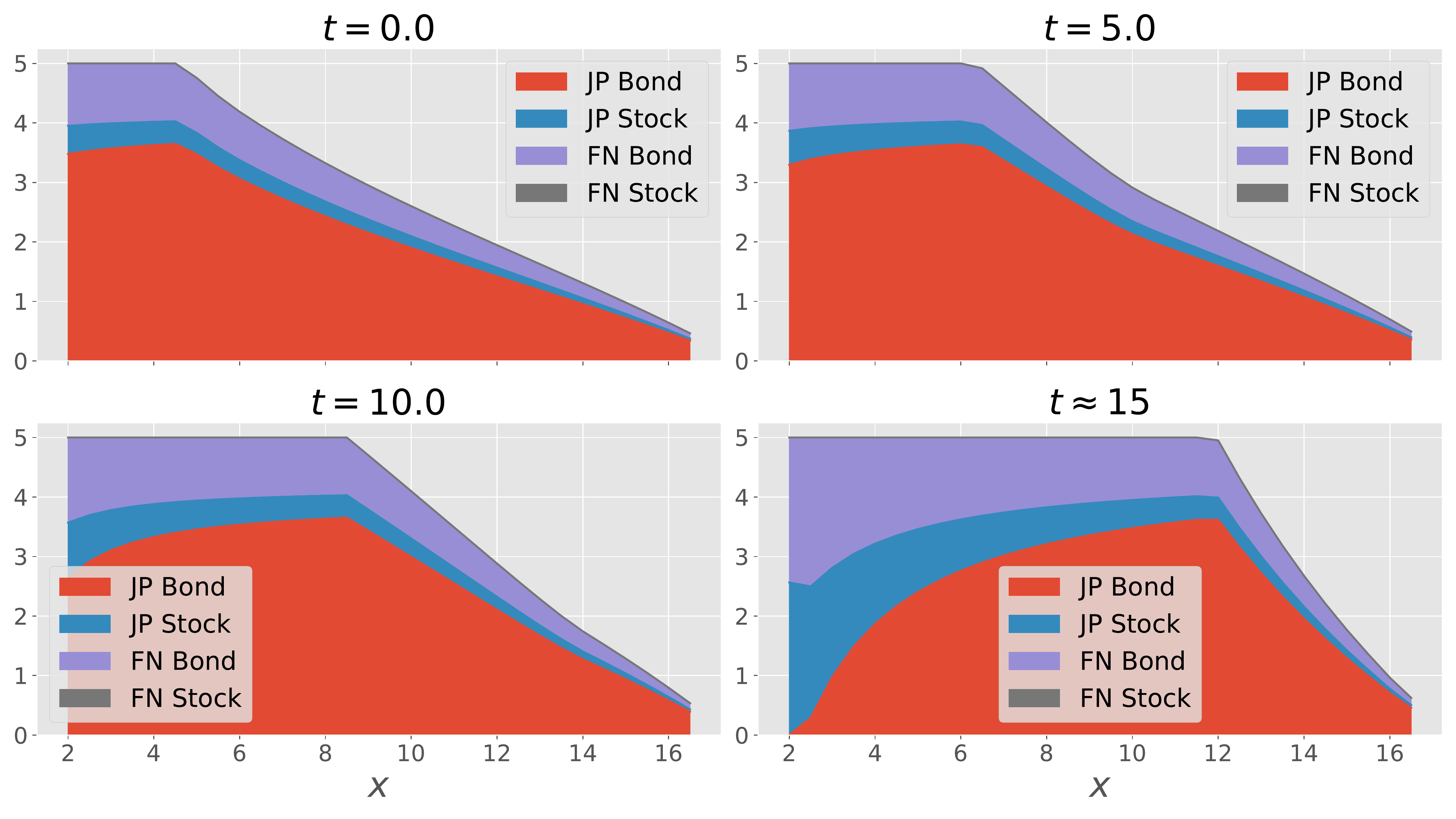}
	\caption{ The optimal portfolio weight for four empirical assets with parameters $\bar{\pi}=5$. }
	\label{fig:emp_stra_pi_5}
\end{figure}

Figure \ref{fig:emp_stra_pi_5} shows the optimal portfolio weight in the case of $\bar{\pi}=5$, which represents the proxy of the unconstrained leverage.
The ingredient of the strategy is simple: adjusting the exposure of the investor's wealth level and keeping the same proportion of weights among the invested assets.
This implication is consistent with previous studies.

Figure \ref{fig:emp_stra_pi_1} shows the optimal portfolio under the leverage constraint, which is our case of interest.
It is a little more complex compared with the unconstrained one.
The strategy consists of three stages, depending on investor's wealth and the remaining time.
We list the stages in ascending order of investor wealth $x$:
\begin{description}
	\item[High risk stage] 
		We invest intensively in the index of Japanese stocks with expected return $b^2=4.8\%$.
	\item[Middle risk stage]
		The main assets are the indices of bonds, while the index of Japanese stocks is invested in complementarily.
		The foreign bonds index plays the central role if investor's wealth level is low 
	\item[Low risk stage] 
		We shrink the exposure while keeping the proportion of the weights among the invested assets.
		The strategy in this stage coincides with the unconstrained case.
\end{description}

\begin{figure}[H]
	\centering
	\includegraphics[width=35em]{./graph/gpif_strategy_pi=1.pdf}
	\caption{ The optimal portfolio weight for four empirical assets with parameters $\bar{\pi}=1$.}
	\label{fig:emp_stra_pi_1}
\end{figure} 

\begin{figure}[H]
	\centering
	\includegraphics[width=33em]{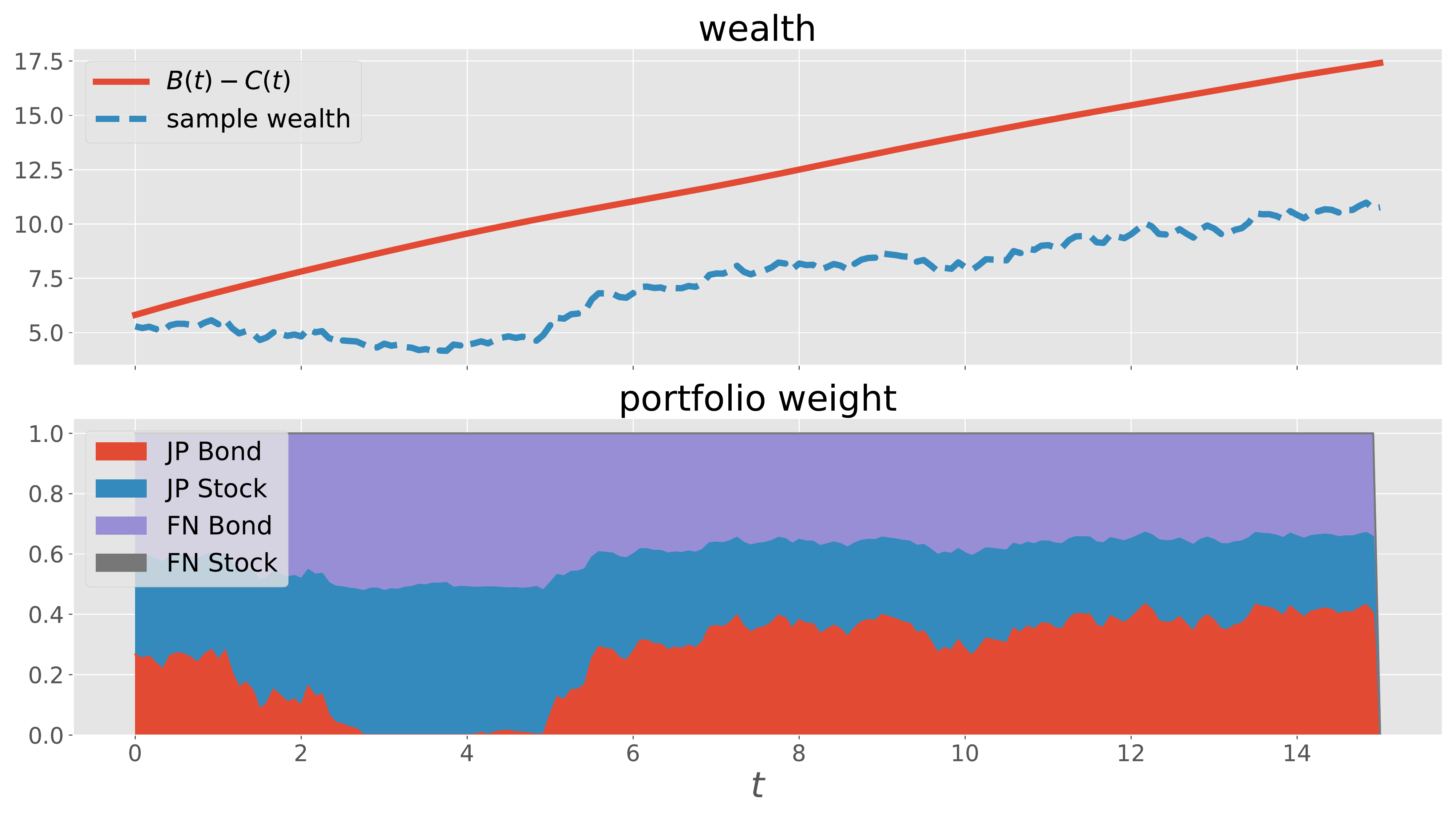}
	\caption{
		A sample path with the empirical data.
	}
	\label{fig:emp_sample_path}
\end{figure}

To explain these stages, see Figure \ref{fig:emp_sample_path} with the sample path of the investment simulation.
The investor is in the high-risk stage from $t\approx 3$ to $t \approx 5$ and in the middle-risk stage in the remaining period.
Under the leverage constraint $\bar{\pi} =1$, the low-risk stage rarely appears.

\subsection*{Simulation results}
\begin{figure}[H]
	\centering
	\includegraphics[width=33em]{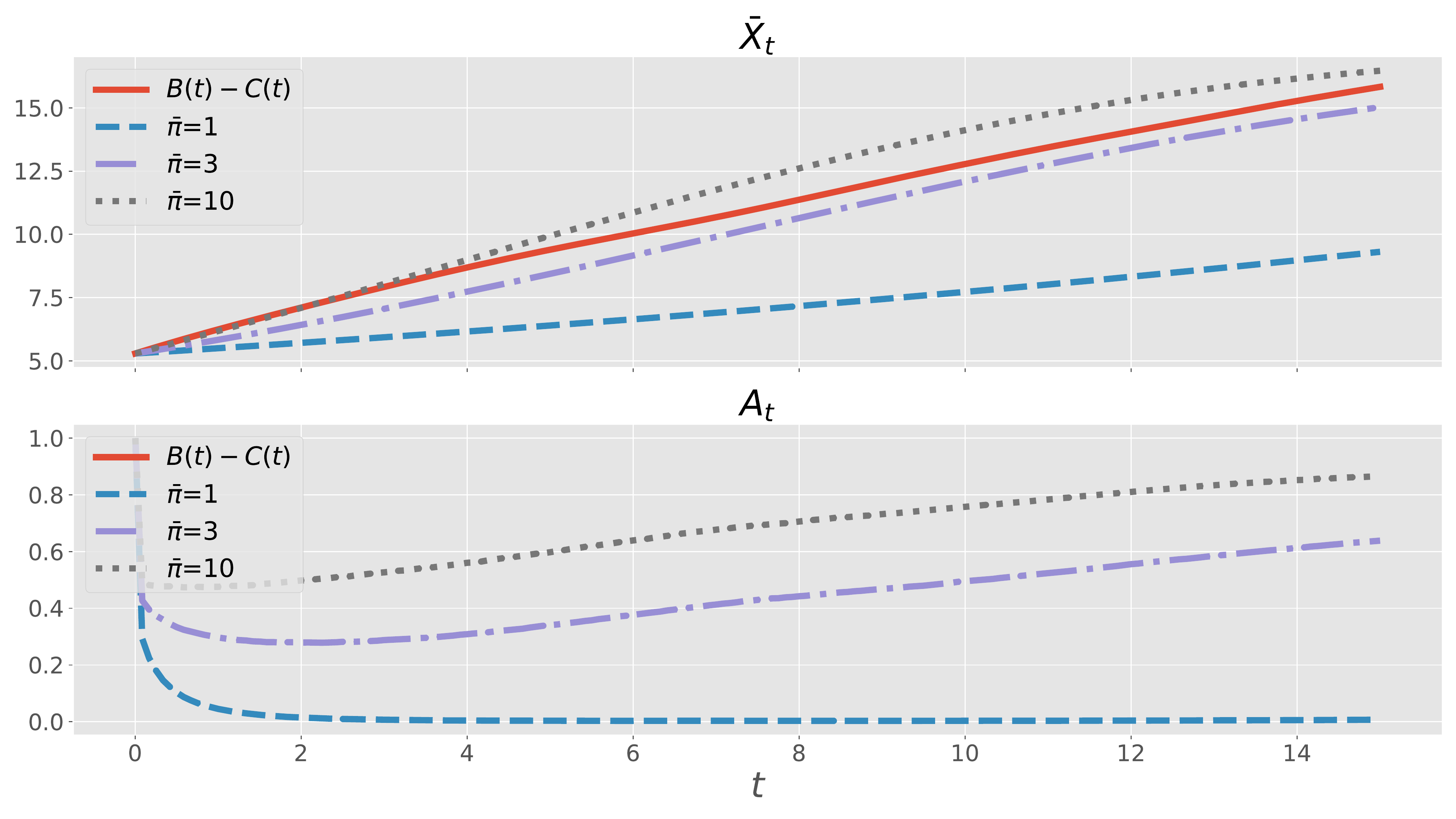}
	\caption{ Simulation results $\bar{X}_t$ and $A_t$ with the empirical data. }
	\label{fig:emp_sim_res}
\end{figure}

Figure \ref{fig:emp_sim_res} shows the simulation results $\bar{X}_t$ and $A_t$ with the empirical data.
Table \ref{tb:emp_comp_stat} indicates the comparison of the statistics with various leverage levels.
In the case of $\bar{\pi}=10$, it appears to be appropriately tracking the target wealth $B(t)-C(t)$
 because the achievement rate $A_T=87\%$ and the mean wealth $\bar{X}_t$ exceeds $B(t)-C(t)$ overall.
A remarkable point is the latter, since the opposite result was obtained in previous studies.
The margin equipped to calculate the optimal portfolio contributes to this matter.

\begin{table}[H]
	\centering
	\caption{Comparison of the statistics with various leverage levels.}
	\label{tb:emp_comp_stat}
\begin{tabular}{llllll}
	\toprule
	$\bar{\pi}$ & $B(T)-C(T)$ & $\bar{X}_T$ & $A_{T/2}$ & $A_T$ &  $P_T$ \\
	\midrule
	1 &             &        9.31 &       0\% &   1\% &   4.68 \\
	2 &             &       13.08 &      16\% &  33\% &   6.76 \\
	3 &             &       15.03 &      43\% &  64\% &   7.71 \\
	4 &       15.83 &       15.77 &      57\% &  76\% &   8.42 \\
	5 &             &       16.09 &      63\% &  81\% &   9.29 \\
	7 &             &       16.34 &      67\% &  85\% &  11.82 \\
	10 &             &       16.47 &      69\% &  87\% &  13.09 \\
	\bottomrule
\end{tabular}
\end{table}

If we prohibit leverage, investment performance cannot reach the desired level.
The achievement rate is almost $0\%$, and thus, we must relax our investment conditions: 
 choose another investment asset,
 reduce the target wealth,
 or weaken the leverage constraint.
If we decide to weaken the leverage constraint, the choice of its value is subjective.
For instance, selecting $\bar{\pi}=5$ appears to be a possible choice because the achievement rate reaches 80\%.
It is unclear whether choosing the leverage level $\bar{\pi}=3$ is correct.
On the one hand, the mean wealth appears to be close to the target wealth.
On the other hand, the achievement rate $A_T = 64\%$ pales in comparison to the cases where $\bar{\pi}\geq 4$.

\section{Summary}
We investigated the continuous-time portfolio optimization problem with
 (i) a no-short selling constraint;
 (ii) a leverage constraint, that is, an upper limit for the sum of portfolio weights;
 and (iii) a performance criterion based on the LMSE between our wealth and a predetermined target wealth level.
To accomplish this aim, we advanced the model and the methodology discussed in the previous work by \autocite{Ieda2014}.

We reformulated the model with the leverage constraint in a natural way.
Furthermore, the dimension of the state variable was decreased from two to one, and thus, the computational load was notably reduced.
Moreover, the boundary condition of the corresponding HJB equation was obtained explicitly.
Although the numerical method employed here can be applied without boundary conditions, it improved the accuracy of the numerical solution.

We employed the kernel collocation method instead of the quadratic approximation 
  because it only provides a proxy for the solution under the leverage constraint.
As shown in the numerical results section, the straightforward implementation did not work well;
 the approximated derivatives behaved unstably because of the discontinuity of the second derivative of the terminal condition at the boundary point.
We circumvented this matter by locating the extra sample points outside the boundary.

Two numerical results were found: one using artificial data and the other using empirical data from Japanese organizations.
A key implication of the result using artificial data is that
 the obtained strategy did not achieve the target wealth level close to the terminal time,
 even though it was achieved in the running period; the feature of LMSE caused it.
We addressed this issue by using the incentive to exceed the target level, such as taking a longer terminal time or introducing the margin.

The optimal strategy with the leverage constraint for the empirical data was divided into three stages,
 depending on investor's wealth and the remaining time:
(i) a high-risk stage in which we invested in the index of Japanese stocks intensively;
(ii) a middle-risk stage in which the main assets were the indices of bonds and the index of Japanese stocks was invested in complementarily; and
(iii) a low-risk stage in which we shrinked the exposure while keeping the proportion of the weights among the invested assets.
However, we were not able to achieve the required wealth level without leverage.
Therefore, leverage is necessary, and we must choose the leverage level with which we can compromise if we do not change the present setting.

\section*{Acknowledgment}
This work is supported by JSPS Grant-in-Aid for Young Scientists(Start-up) Grant Number JP20K22130.

\printbibliography 

\end{document}